\DocumentMetadata{
  lang        = en,
  pdfversion  = 2.0,
  pdfstandard = A-2b,
  testphase   = 
    {phase-III,
     title,
     table, math}  
}

\documentclass[examplefnt,biber]{nowfnt} 

\usepackage{amsmath,amsfonts,bm}
\usepackage{array}
\usepackage{booktabs}
\usepackage{multirow}
\usepackage{caption}
\usepackage{url}
\usepackage{rotating}
\usepackage[utf8]{inputenc}
\usepackage{graphicx}

\usepackage{tagpdf}
\tagpdfsetup{
  activate-all,
  uncompress,
  para/tagging = true,
}
\tagpdfparaOn

\hypersetup{
    colorlinks = true,
    linkcolor = blue,
    urlcolor = magenta,
    citecolor = blue,
    anchorcolor = blue
}

\providecommand{\gz}[1]{\textcolor{black}{{#1}}}

\title{Audio Generation Through Score-Based Generative Modeling: Design Principles and Implementation}

\maintitleauthorlist{
Ge Zhu \\
Department of Electrical and Computer Engineering \\
University of Rochester \\
ge.zhu@rochester.edu
\and
Yutong Wen \\
Department of Electrical and Computer Engineering \\
University of Rochester \\
ywen6@u.rochester.edu
\and
Zhiyao Duan \\
Department of Electrical and Computer Engineering \\ 
University of Rochester \\
zhiyao.duan@rochester.edu
}

\issuesetup
{%
 copyrightowner={G.~Zhu, Y.~Wen and Z.~Duan},
 volume        = xx,
 issue         = xx,
 pubyear       = 2024,
 isbn          = xxx-x-xxxxx-xxx-x,
 eisbn         = xxx-x-xxxxx-xxx-x,
 doi           = 10.1561/XXXXXXXXX,
 firstpage     = 1, 
 lastpage      = 18
 }

\author[1]{Zhu,Ge}
\author[2]{Wen,Yutong}
\author[3]{Duan,Zhiyao}

\affil[1]{Department of ECE, University of Rochester; ge.zhu@rochester.edu}
\affil[2]{Department of ECE, University of Rochester; ywen6@u.rochester.edu}
\affil[2]{Department of ECE, University of Rochester; zhiyao.duan@rochester.edu}

\articledatabox{\nowfntstandardcitation}

\begin{document}
\makeabstracttitle

\begin{abstract}
Diffusion models have emerged as powerful deep generative techniques, producing high-quality and diverse samples in applications in various domains including audio.
\gz{While existing reviews provide overviews, there remains limited in-depth discussion of these specific design choices.}
The audio diffusion model literature also lacks principled guidance for the implementation of these design choices and their comparisons for different applications.
This survey provides a comprehensive review of diffusion model design with an emphasis on design principles for quality improvement and conditioning for audio applications. 
We adopt the score modeling perspective as a unifying framework that accommodates various interpretations, including recent approaches like flow matching.  
We systematically examine the training and sampling procedures of diffusion models, and audio applications through different conditioning mechanisms.
\gz{To provide an integrated, unified codebase and to promote reproducible research and rapid prototyping, we introduce an open-source codebase (\url{https://github.com/gzhu06/AudioDiffuser}) that implements our reviewed framework for various audio applications.} 
We demonstrate its capabilities through three case studies: audio generation, speech enhancement, and text-to-speech synthesis, with benchmark evaluations on standard datasets.

\end{abstract}

\chapter{Introduction}
\label{c-intro} 


Diffusion models have emerged as a powerful class of generative models, achieving state-of-the-art performance in various domains~\citep{yang2023diffusion}. 
There are two primary formulations for diffusion models: denoising diffusion probabilistic models (DDPMs)~\citep{ho2020denoising}, and score-based generative models (SGMs)~\citep{song2020score}.
DDPMs, which originate from non-equilibrium thermal dynamics~\citep{sohl2015deep}, construct a forward Markov chain that progressively adds noise to the data, together with a reverse chain that learns to transform the noise back to the data using neural network-parameterized transition kernels.
SGMs, on the other hand, originating from score matching~\citep{hyvarinen2005estimation}, use neural networks to learn the Stein score function of the data distribution through noise perturbation~\citep{liu2016kernelized}.

Although initially developed as separate approaches, these frameworks were elegantly unified through stochastic differential equations (SDEs) in \gz{score-based framework}~\citep{song2020score}, which demonstrated that both can be described by the same underlying SDE with different drift and diffusion terms, providing a more comprehensive theoretical foundation.
\gz{Major subsequent developments include the elucidated diffusion model (EDM) framework~\citep{karras2022elucidating}, which systematically separates the design space into independent components: sampling processes, training objectives, and network preconditioning and derives principled scaling functions that improve training stability and sampling efficiency.}
Further innovations include efficient sampling techniques~\citep{lu2022dpm}, latent diffusion models~\citep{rombach2022stable}, and classifier-free guidance~\citep{ho2022classifier}, significantly improving generation quality and computational efficiency.

The diffusion model approach applies naturally to audio signals due to its distribution-agnostic nature.
These models have demonstrated superior performance in audio applications including vocoding~\citep{kong2020diffwave}, text-to-speech synthesis~\citep{popov2021grad}, music generation~\citep{evans2024stable}, speech enhancement~\citep{richter2023speech}, and source separation~\citep{yuan2024flowsep}.
Their success stems from two key advantages: the capacity to generate high-fidelity audio and the flexibility for conditional generation.

There have been reviews of diffusion models, each with different focus areas. 
\citet{yang2023diffusion} provide a broad overview of diffusion models for generative tasks, while~\citet{croitoru2023diffusion} concentrate specifically on computer vision applications. 
\gz{In the audio domain, \citet{zhang2023survey} examine diffusion models broadly, while \citet{lemercier2024diffusion} provide a more recent and focused review on audio restoration tasks including speech enhancement and music restoration.}
\gz{While these comprehensive reviews provide excellent coverage of diffusion model theory and applications, our work complements them by providing an integrated perspective that connects mathematical formulations to practical design choices in real-world applications.}

\begin{figure*}[!t]
\centering
\includegraphics[width=1\columnwidth]{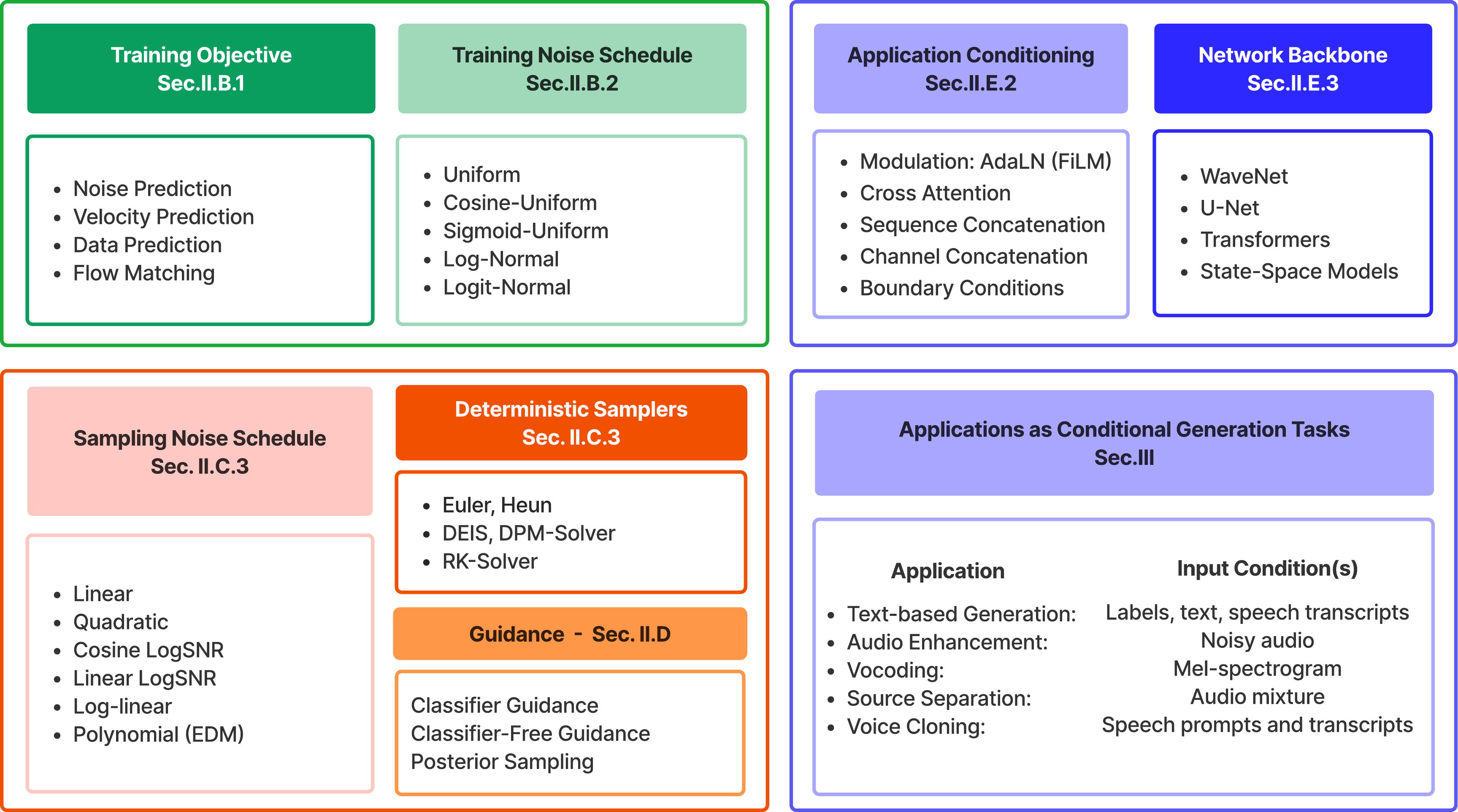}
\caption{
\gz{Design space of diffusion-based generative models for audio applications. 
The diagram illustrates the key modular components for diffusion models, organized into three categories: (1) training components, including training objectives (noise/velocity/data prediction, flow matching) and noise schedules (uniform, cosine-uniform, sigmoid-uniform, etc.); (2) sampling components, including deterministic samplers (Euler, Heun, DEIS, DPM-Solver, RK-Solver), sampling noise schedules, and guidance mechanisms (classifier guidance, classifier-free guidance, posterior sampling); (3) model architecture, including conditioning mechanisms (modulation, cross-attention, concatenation) and network backbones (WaveNet, U-Net, Transformers, State-Space Models). This modular design space has been systematically analyzed by frameworks like EDM~\citep{karras2022elucidating} and extends to recent flow matching approaches. The interchangeability of these components enables flexible experimentation across diverse audio generation tasks, from text-to-audio to voice cloning.}
}
\label{fig:diffsummary}
\end{figure*}

In this paper, we review diffusion models through the lens of the EDM~\citep{karras2022elucidating}, a principled approach within the SGM family.
A diagram for the review is shown in Fig.~\ref{fig:diffsummary}.
\gz{Compared to prior diffusion formulations such as DDPM~\citep{ho2020denoising}, EDM improves both sample quality and convergence speed through systematic optimization of noise schedules, network preconditioning, and loss weighting. 
While EDM has primarily demonstrated strong performance on image generation, we adopt it to the audio domain and validate that its principled training and sampling practices transfer effectively.}


This paper makes three primary contributions:
\begin{itemize}
    \item We provide a systematic review of diffusion models using EDM, emphasizing design principles that impact generation quality and  conditioning flexibility for audio applications;
    \item We present a comprehensive review of representative audio applications utilizing diffusion models, categorizing them by task type and highlighting architectural choices and conditioning mechanisms;
    \item We introduce an open-source codebase that implements the key components discussed in our theoretical exposition, supporting various audio applications and facilitating reproducibility and further research.
\end{itemize}
By bridging theoretical understanding with practical implementation guidance, this paper aims to equip audio researchers with the knowledge and tools necessary to effectively leverage diffusion models for their specific applications.



\chapter{Score-based Generative Models}
\label{theory}

\label{sec:main}
In this section, we examine diffusion models from the perspective of SGMs, with a primary focus on the EDM framework~\citep{karras2022elucidating} as an organizational lens for practical implementation.
\gz{Our goal is not to provide a comprehensive theoretical survey of all diffusion formulations, but rather to offer actionable guidance for practitioners building audio generation systems.
EDM is particularly well-suited for this purpose as it systematically decomposes the design space of diffusion models into modular components including noise schedules, preconditioning schemes, training loss weighting, and sampling methods.  
By explicitly tabulating these components, EDM enables principled analysis and optimization of each element independently. }
In line with this approach, we discuss each diffusion component individually.
Given that diffusion processes can model data distribution across any modality, we explore general design principles for training denoising networks that remain effective regardless of the underlying data type.
We then investigate efficient sampling techniques and the application of guidance along the sampling path for ordinary differential equation (ODE) solvers. 
Finally, we discuss network architecture design choices and explore conditioning methods specifically for audio applications, demonstrating how diffusion models can be effectively applied to this domain.

\section{Formulation of Score-based Generative Models}
\label{method:overview}
\gz{In SGMs, the goal is to estimate the score function $\nabla_{\mathbf{x}} \log p_{\text{data}}(\mathbf{x})$, which represents the gradient of the log-probability density of the data distribution. 
Since we lack explicit access to this score function during training, denoising score matching (DSM)~\citep{vincent2011connection} circumvents this by optimizing over paired clean and corrupted data samples, where corruption is induced by Gaussian perturbations. 
Specifically, the model learns to estimate $\nabla_{\mathbf{x}} \log q(\mathbf{x}; \sigma^2\mathbf{I})$, where $q(\mathbf{x}; \sigma^2\mathbf{I})$ represents the distribution obtained by convolving the data distribution with Gaussian noise of variance $\sigma^2$, and $\sigma$ is the standard deviation of this perturbation. 
By training on data perturbed at multiple predefined noise levels, SGMs can effectively model arbitrary distributions.}

A key advantage of this gradient-based formulation is that the score function remains independent of the typically intractable normalization constant of the density function, which makes it easier to evaluate~\citep{hyvarinen2005estimation}.
Intuitively, the score function can be viewed as a vector field that indicates the directions along which the probability density function increases most rapidly at a given noise level $\sigma$.
The perturbation noise can be continuous or discrete spanning a range such that at the maximum noise level $\sigma_{\max}$, $q(\mathbf{x};\sigma^2_{\max}\mathbf{I})$ is indistinguishable from Gaussian distribution and at the minimum level $\sigma_{\min}$, $q(\mathbf{x};\sigma^2_{\min}\mathbf{I})$ approximates the \gz{original} data distribution.

In this paper, we opt for a continuous noise schedule where the noise level $\sigma(t)$ varies as a function of time parameter $t$.
\gz{For such continuous noise levels, SGMs can be described as stochastic differential equations (SDEs)~\citep{song2020score,yang2023diffusion}. 
Specifically, the forward diffusion process is defined by the following SDE that gradually perturbs clean data with noise:}
\begin{equation}
d{\bf{x}}={\bf{f}}({\bf{x}},t)dt + g(t)d{\bf{w}},
\label{eq:sde}
\end{equation}
where ${\bf{f}}({\bf{x}},t)$ and $g(t)$ are \textit{drift} and \textit{diffusion} functions of the SDE, respectively, and $\bf{w}$ denotes a standard Wiener process. 
The commonly used ${\bf{f}}({\bf{\cdot}})$ takes the form 
\begin{equation}
    \mathbf{f}(\mathbf{x},t)=f(t){\mathbf{x}}
    \label{eq:drift},
\end{equation}
where $f(\cdot):\mathbb{R}\to \mathbb{R}$. 
This formulation yields a forward Gaussian noise perturbation kernel~\citep{song2020score,karras2022elucidating}:
\begin{equation}
    q(\mathbf{x}(t)|\mathbf{x}(0))
=\mathcal{N}(s(t)\mathbf{x}(0),s(t)^2\sigma^2(t)\mathbf{I}),
\label{eq:noising}
\end{equation}
where $s(t)$ and $\sigma(t)$ are functions of $f(t)$ and $g(t)$ in Eqs.~\eqref{eq:sde} and~\eqref{eq:drift}~\citep{karras2022elucidating}.
Specifically, $s(t)=\exp(\int_0^t f(\tau)d\tau)$ represents the time-dependent scaling factor, while $\sigma^2(t)=\int_0^t \frac{g(\tau)^2}{s(\tau)^2}d\tau$ determines the noise variance. 
Together, these functions define how the original data is progressively scaled and perturbed during the diffusion process.
The DSM training objective optimizes over these paired clean and corrupted samples. Ignoring noise-dependent weighting, the objective can be formulated as~\citep{vincent2011connection,song2020score,ho2020denoising}:
\begin{equation}
\underset{\mathbf{s}_{\theta}}{\text{min}}\mathbb{E}_{t,\mathbf{x}(0)} \mathbb{E}_{\mathbf{x}(t)|\mathbf{x}(0)}||\mathbf{s}_{\theta}(\mathbf{x}(t),\sigma(t))-\nabla_{\mathbf{x}(t)} \log q(\mathbf{x}(t)|\mathbf{x}(0))||^2_2,
\label{eq:dsm}
\end{equation}
where $q$ follows the Gaussian perturbation kernel defined in the noising process Eq.~\eqref{eq:noising}, and $\mathbf{s}_{\theta}(\mathbf{x}(t),\sigma(t))$ is the score predicted with a neural network parameterized by $\theta$ to be estimated.
Due to the Gaussian expression of $q$, the training target in Eq.~\eqref{eq:dsm} can be analytically expressed as:
\begin{equation}
    \nabla_{\mathbf{x}(t)} \log q(\mathbf{x}(t)|\mathbf{x}(0))=-\frac{\mathbf{x}(t)-s(t)\mathbf{x}(0)}{s(t)^2\sigma^2(t)}.
    \label{eq:gaussian}
\end{equation} 
In practice, with sufficient data and model capacity, the score network trained by minimizing Eq.\eqref{eq:dsm} \gz{on finite training samples is expected to generalize and closely approximate the score function of the entire perturbed data distribution} $\nabla_{\mathbf{x}} \log q(\mathbf{x}; \sigma^2\mathbf{I})$~\citep{song2020score,vincent2011connection}.

The score networks $\mathbf{s}_{\theta}(\mathbf{x},\sigma)$ in the DSM objective can be reparameterized differently in practice when substituting Eq.~\eqref{eq:gaussian} into Eq.~\eqref{eq:dsm}. 
\gz{Common alternatives include predicting the added noise, predicting the clean data directly (denoising), or predicting a velocity that combines both~\citep{kingma2024understanding}. 
In this work, we adopt the denoising function reparameterization~\citep{vincent2011connection,karras2022elucidating}, expressed as:}
\begin{equation}
    \mathbf{s}_{\theta}(\mathbf{x},\sigma)=\frac{D_\theta(\mathbf{x}, \sigma)-\mathbf{x}}{\sigma^2},
    \label{eq:sdsm}
\end{equation}
where $D_\theta(\mathbf{x}, \sigma)$ represents a neural network that learns to estimate the conditional expectation of the clean signal given noisy input $\mathbf{x}$ and noising process variance $\sigma^2$.
This conditional expectation arises because a noisy observation could have been generated from different clean signals due to the randomness of the noising process; the denoiser's goal is thus to ensure that the expectation of the inferred clean signal is close to the true clean signal.

To generate new data, the reversed noise perturbation process can be applied.
For the diffusion process defined in Eq.~\eqref{eq:sde}, its reverse process is also a diffusion process running backward in time~\citep{anderson1982reverse,song2020score}:
\begin{equation}
d{\bf{x}}=[{\bf{f}}({\bf{x}},t) -g(t)^2\nabla_{\mathbf{x}} \log q(\mathbf{x}; \sigma(t))]dt +g(t)d{\bf{\bar{w}}},
\label{eq:rsde}
\end{equation}
where ${\bf{\bar{w}}}$ is a standard Wiener process in reverse time.
Numerical SDE solvers can then be applied to Eq.~\eqref{eq:rsde} using estimated score functions $\mathbf{s}_{\theta}(\mathbf{x}(t),\sigma(t))$.
A remarkable property of the reverse SDE is the existence of a corresponding deterministic process whose trajectories share the same $q(\mathbf{x}; \sigma(t))$ as the original stochastic process~\citep{song2020score}.
This deterministic process, termed as \textit{probability flow ordinary differential equation} (PF ODE), can be written as:
\begin{equation}
d{\bf{x}}=[{\bf{f}}({\bf{x}},t) - \frac{1}{2}g(t)^2\nabla_{\mathbf{x}} \log q(\mathbf{x}; \sigma(t))]dt.
\label{eq:ode}
\end{equation}
This allows the use of deterministic ODE solvers for data generation. 
\gz{The performance of deterministic samplers versus stochastic ones depends on the specific application and requires further investigation. 
Prior work in generative modeling has shown mixed results: \citet{song2020score} reported advantages for deterministic samplers on simpler datasets, while \citet{karras2022elucidating} found that stochastic sampling performs better on more diverse datasets, suggesting that dataset complexity may influence the optimal sampler choice.}
In this paper, we focus on ODE-based samplers, as they offer advantages over their SDE counterparts in terms of sampling efficiency.
\gz{The step size in SDE solvers is typically limited by the statistical properties of the driving Wiener process.
Because Brownian increments scale as $N(0,\Delta t)$, the pathwise error of standard discretizations such as the Euler–Maruyama scheme grows as $O(\sqrt{\Delta t})$~\citep{kloeden1992numerical, higham2001algorithmic}.
Consequently, reducing this stochastic error requires quadratically smaller step sizes, which constrains the practical discretization step.
Deterministic ODE solvers, on the other hand, are not subject to such stochastic scaling and can often achieve similar accuracy with larger steps.}
\gz{Another benefit of ODE-based samplers is that they provide deterministic, approximately invertible mappings between data and latent spaces, enabling applications such as editing and interpolation~\citep{song2021denoising,song2020score}. 
The approximate invertibility arises from the deterministic nature of the ODE trajectory, though numerical discretization introduces small errors.}


\gz{While our presentation focuses on the score-based diffusion framework, several alternative formulations have emerged that offer distinct theoretical and practical advantages. 
Flow matching~\citep{lipman2023flow,liu2023flow} provides a simulation-free training paradigm that directly regresses vector fields along interpolating paths (e.g., linear or optimal transport paths) between noise and data distributions.  
This formulation offers computational advantages through simplified training objectives and can achieve faster sampling by learning straighter trajectories between distributions~\citep{liu2023flow}.
Diffusion bridges~\citep{shi2023diffusion,de2021diffusion,zhou2024denoising} extend the standard diffusion framework to enable conditioning on both endpoints of the diffusion process, effectively learning the distribution $q(\mathbf{x}(t)|\mathbf{x}(0), \mathbf{x}(T))$ rather than just $q(\mathbf{x}(t)|\mathbf{x}(0))$.
This is achieved through Doob's h-transform~\citep{rogers2000diffusions}, which modifies the drift term of the forward SDE in Eq.~\eqref{eq:sde} by incorporating an additional term that guides the process toward a specified endpoint $\mathbf{x}(T)$. 
This formulation naturally facilitates tasks requiring controllable transformations between specified states, such as data-to-data translation and conditional generation where both source and target domains are constrained.}

To summarize the overall framework, the training process involves optimizing denoising network $D_\theta(\mathbf{x}, \sigma)$ to predict the expected clean signal given noisy inputs through the score matching objective in Eq.~\eqref{eq:dsm}.
For generation, we leverage the trained network to compute the score function, which enables solving the probability flow ODE in Eq.~\eqref{eq:ode}.
This generation process starts with Gaussian noise samples $(t=T)$ and deterministically transforms them to the data distribution $(t=0)$.
\gz{Deterministic sampling can offer practical advantages for applications requiring reproducible outputs or latent space manipulation.}

\section{Training}
\label{sec:train}
\subsection{Reparameterized and Weighted Training Objective}
\label{sec:train_obj}
To reduce score function fitting error, training objective reparameterization~\citep{salimans2022progressive,karras2022elucidating,karras2023analyzing} and loss weighting~\citep{vahdat2021score,karras2022elucidating,kingma2024understanding} are commonly used effective methods. 

Within the denoising function formulation, the vanilla training objective Eq.~\eqref{eq:dsm} can be expressed as:
\begin{equation}
    \mathbb{E}_{\sigma} \mathbb{E}_{\mathbf{x}(t)|\mathbf{x}(0)}[\lambda(\sigma)||D_{\theta}(\mathbf{x}(t),\sigma(t))-\mathbf{x}(0)||^2_2].
    \label{eq:trainobj}
\end{equation}
The EDM framework, in particular, preconditions neural networks to facilitate objective prediction, addressing challenges in both high- and low-noise scenarios.
EDM represents $D_\theta(\mathbf{x},\sigma)$ as a combination of weighted data skip connection and a neural network $F_\theta$ that predicts noise components:
\begin{equation}
D_\theta(\mathbf{x},\sigma)=c_{\text{skip}}(\sigma)\mathbf{x} + c_{\text{out}}(\sigma)F_\theta(c_{\text{in}}(\sigma)\mathbf{x}(\sigma), c_{\text{noise}}(\sigma)).
    \label{eq:edmobj}
\end{equation}
This noise-level-dependent scaling enables the neural network to adapt its prediction objective based on noise magnitude, effectively transitioning between denoising at low noise levels and high noise levels.
Additionally, this scaling provides a configuration that satisfies the boundary conditions required for consistency models~\citep{song2023consistency}.
This reparameterization serves multiple purposes: (1) The coefficients $c_{\text{in}}$ and $c_{\text{out}}$ are used to normalize the magnitudes of the network input and training targets, respectively, ensuring a unit variance.
The benefit of using input normalization $c_{\text{in}}$ is also observed in~\citet{chen2023importance}.
(2) The coefficient $c_{\text{skip}}$ is introduced to balance the relative contributions of $\mathbf{x}$ and $F_\theta(\mathbf{x},\sigma)$ to the final output. 
This control mechanism is designed to reduce the prediction error that has been amplified by $c_{\text{out}}$.
(3) The noise parameter $\sigma$, which conditions the network, is transformed using the function $c_{\text{noise}}(\sigma)$ to represent noise levels on a logarithmic scale.
This logarithmic transformation of the noise level input allows for better handling of different noise scales, a similar technique used in Imagen~\citep{saharia2022photorealistic}.
Training objective reparameterization can also reduce the score sampling discretization error.
In the commonly used $\epsilon$-prediction parameterization, Salimans and Ho~\citep{salimans2022progressive} proposed the velocity prediction objective ($\bf{v}$-prediction). 
This method replaces the network output with a mixture of $\epsilon$ and $\mathbf{x}$, addressing instabilities that can occur in $\epsilon$-prediction at high perturbation noise level, particularly in few-step sampling scenarios.

The loss weighting term, on the other hand, aims to normalize the loss magnitude for each noise level, standardize gradient magnitudes across different noise levels, and prevent any particular noise level from dominating the optimization process. 
To achieve this goal under the preconditioning scheme defined earlier, the weighting function $\lambda(\sigma)$, defined in Eq.~\eqref{eq:trainobj}, is initially set to $1/c_{\text{out}}$.
However, as training progresses, this initial setting is found suboptimal~\citep{karras2023analyzing}: The magnitude of gradient feedback begins to vary between noise levels, leading to an uncontrolled re-weighting of their relative contributions. 
Furthermore, it has been demonstrated that the optimal weight gradients for different noise levels often conflict with one another, resulting in slow convergence~\citep{hang2023efficient}.
To address these issues, multi-task based approaches have been proposed in~\citet{hang2023efficient,karras2023analyzing}. 
These methods treat the training of each noise level as an individual task and apply a noise-level-dependent weighting function to the loss. 
This approach allows for more precise control over the contribution of each noise level throughout the training process, potentially leading to faster convergence and improved model performance.


\subsection{Noise Schedule for Training}
\label{sec:trainnoise}
In \gz{score-based} framework, the functional form of training noise schedule is not necessarily the same as sampling noise schedule.
Naturally, the range of the training noise schedule covers the entire sampling noise schedule, which ensures that the denoising function is well-defined and can effectively estimate denoised data at each sampling noise level, thereby guaranteeing the model's ability to handle all potential noise levels encountered during the sampling process.

Recent research has revealed that not all noise levels contribute equally to the sampling quality during generation~\citep{choi2022perception,raya2024spontaneous}. 
\citet{chen2023importance} and \citet{nichol2021improved} also investigated different training noise schedule strategies on images of different resolutions.
Their findings indicate that shifting training noise schedules towards higher noise levels as image resolution increases yields better generation quality. 
Specifically, cosine schedules, which allocate more timesteps to higher noise regions compared to linear schedules, demonstrated superior performance.
In perception prioritized (P2) diffusion model training, \citet{choi2022perception} suggest that diffusion models learn different types of features at various noise levels: at high noise levels, the model learns coarse features; while at medium noise levels, it captures perceptually rich content; finally, at low noise levels, the model focuses on sharpening the data. 
Based on this hypothesis, they propose to assign higher weights to the loss at noise levels where the model learns perceptually rich content.
An analysis of the fitting error distribution, as described in~\citet{karras2022elucidating}, reveals relative higher errors at both low and high noise levels at convergence. 
This phenomenon likely occurs because at very low noise levels, the model struggles to differentiate between noisy and clean images, while at high noise levels, the model tends to approximate the dataset average rather than to preserve individual target characteristics at convergence. 
As such, training should focus on intermediate noise levels where more substantial error reduction can be achieved.
To address this, EDM applies a log-normal distribution, defined as $\ln \sigma \sim \mathcal{N}(P_{\text{mean}}, P_{\text{std}}^2)$, where $P_{\text{mean}}$ controls the mean noise level and $P_{\text{std}}^2$ represents the variance of the distribution. 
This concentrates sampling density around critical noise levels while spanning multiple orders of magnitude, efficiently allocating gradient updates where most beneficial. 
Empirical observations confirm that these hyperparameters remain robust across various pixel-domain image datasets~\citep{karras2022elucidating}.
For latent embeddings of variational auto-encoders (VAEs), which require higher noise levels for complete data corruption than pixel-space data~\citep{karras2023analyzing,chen2024deconstructing}, an increased $P_{\text{mean}}$ value is necessary.

\section{Sampling}
\label{sec:sampling}
\subsection{Simplification of PF ODE}
\begin{figure}[!t]
\centering
\includegraphics[width=0.7\columnwidth]{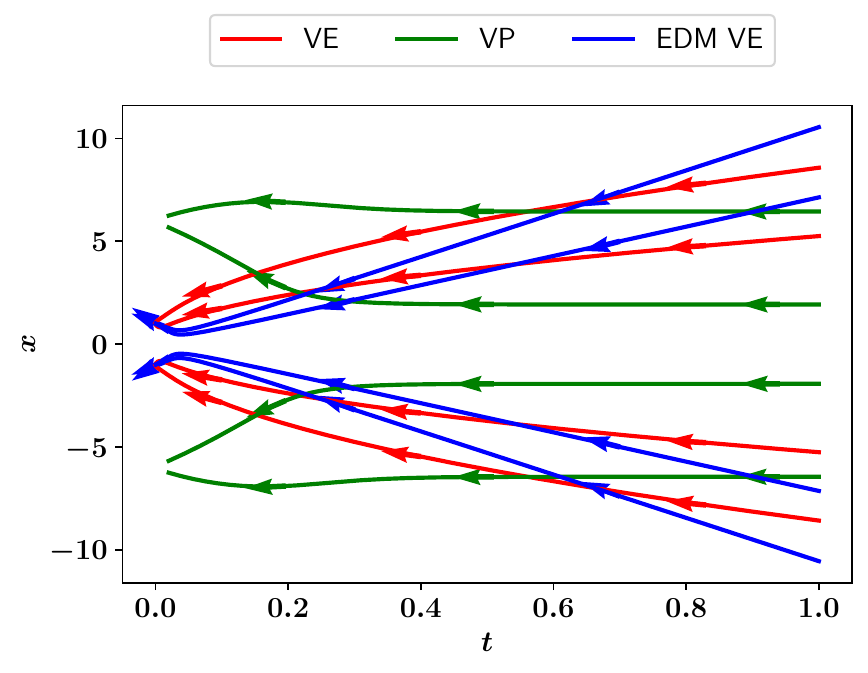}
\caption{Sampling trajectories for approximating a two-peak Dirac distribution, comparing 1D ODE solution curvatures for variance exploding (VE), variance preserving (VP), and EDM VE approaches. 
The horizontal axis $t$ represents the progression from initial noise ($t=1$) to the target distribution ($t=0$), normalized for better illustration. Local gradients for each curve are indicated with arrows.}
\label{fig:sysode}
\end{figure}

During sampling, Eq.~\eqref{eq:ode} is typically used, with functions $\mathbf{f(\cdot)}$ and $g(\cdot)$ often requiring manual designs~\citep{song2020score}. 
However, \citet{karras2022elucidating} argue that $\mathbf{f(\cdot)}$ and $g(\cdot)$ are of limited practical importance, as they only indirectly relate to the marginal distribution $q_{t}(\mathbf{x}(t))$. 
Instead, they propose using the forward noise perturbation parameters $s(t)$ and $\sigma(t)$, as defined in Eq.~\eqref{eq:noising}, for both training and sampling as they directly operate on the distribution.
To do so, the PF ODE can be rewritten as (we refer the readers to Appendix B.2 in~\citet{karras2022elucidating} for more details):
\begin{equation}
    d{\bf{x}}=[\frac{\dot{s}(t)}{s(t)}\mathbf{x}-s(t)^2\dot{\sigma}(t)\sigma(t)\nabla_{\mathbf{x}} \log q(\frac{\mathbf{x}}{s(t)}; \sigma(t))]dt.
\label{eq:edmode}
\end{equation}

The choice of $s(t)$ and $\sigma(t)$ plays a crucial role in determining ODE solution trajectories and the resulting truncation errors when solving Eq.~\eqref{eq:edmode}.
When $s(t)=1$ and $\sigma(t)=\sqrt{t}$, Eq.~\eqref{eq:edmode} transforms into a variance exploding (VE) ODE trajectory. 
In EDM~\citep{karras2022elucidating}, a variant of VE ODEs, the optimal choice for the scale and noise functions is argued to be $s(t)=1$ and $\sigma(t)=t$. 
This specific parametrization reduces the nonlinearity of solution paths, resulting in nearly straight trajectories that can be accurately approximated with fewer integration steps. 
Consequently, this approach achieves minimal truncation errors with a reduced number of NFEs.
To illustrate this concept, Fig.~\ref{fig:sysode} demonstrates sampling trajectories through a toy example of a two-peak Dirac distribution.
This simplified model approximates real-world datasets where data points cluster on low-dimensional manifolds embedded in high-dimensional spaces~\citep{zhang2023gddim,tenenbaum2000global}.
Starting from initial noise, these trajectories follow different ODE solutions parameterized by $s(t)$ and $\sigma(t)$.
Fig.~\ref{fig:sysode} reveals distinct curvature patterns across methods: compared to VP (green) and VE (red), EDM VE (blue) displays near-linear trajectories for a wider range of time from $t=1$ (initial noise) to $t=0$ (target distribution).

From this point forward, we set $s(t)=1$ and $\sigma(t)=t$, following the EDM framework. 
Incorporating these settings and the definition of the denoising function into Eq.~\eqref{eq:edmode}, the probability flow ODE simplifies to:
\begin{equation}
\frac{d{\bf{x}}}{dt}=\frac{\mathbf{x}-D_{\theta}(\mathbf{x}, t)}{t}, 
\label{eq:simedmode}
\end{equation}
where $\frac{d\mathbf{x}}{dt}$ represents the tangent of the solution trajectory, also referred to as ``velocity" or ``flow" in the rectified flow context\citep{liu2023flow}).
This tangent aligns with the vector pointing from an arbitrary point $\mathbf{x}$ towards its denoised counterpart.

The denoising function, $D_{\theta}(\mathbf{x}, t)$, trained with mean squared error (MSE) loss, tends to predict the expected value of the target clean data at given noise levels. 
At high noise levels (larger t), the trajectory tangent points towards the global mean of the data distribution, while at low noise levels (smaller t), it directs towards the local mean of the data manifold~\citep{karras2022elucidating,dieleman2023perspectives}, also evidenced by Fig.\ref{fig:sysode}.

Through the equivalence between the denoising function and perturbation noise, we can reformulate the original ODE.
Specifically, the score function relates to the perturbation noise via 
$\mathbf{s}_{\theta}(\mathbf{x},\sigma)=-\epsilon_{\theta}(\mathbf{x},\sigma)/\sigma$.
By combining this relationship with the denoising function definition in Eq.~\eqref{eq:sdsm} and setting $\sigma(t)=t$, the ODE simplifies to~\citep{zhang2023fast}:
\begin{equation}
\frac{d{\bf{x}}}{dt}=\epsilon_{\theta}(\mathbf{x},t),
\label{eq:noiseode}
\end{equation}
This ODE is shared by denoising diffusion implicit models (DDIMs)~\citep{song2021denoising} and the backward ODE in the flow model context~\citep{liu2023flow,lipman2023flow} when the initial value distribution and the endpoint distribution match those of the PF-ODEs.

\subsection{Noise Schedule for Sampling}
With ODE defined in Eq.~\eqref{eq:simedmode} or Eq.~\eqref{eq:noiseode}, different discretizations of the time steps (also known as noise schedule) and integration schemes can be applied. 
The discretized time steps are a monotonically decreasing sequence $\{t_n\}_{n=0}^N$, where $\sigma_{\max}=t_0>t_1 >...> t_N=0$. 
Commonly used sampling noise schedule as a function of time steps include cosine~\citep{nichol2021improved}, linear~\citep{ho2020denoising}, quadratic~\citep{song2021denoising}, sigmoid~\citep{jabri23scalable}, log-linear~\citep{lu2022dpm}, and polynomial~\citep{karras2022elucidating}.

The choice of noise schedules influences sample quality and efficiency. 
However, few prior works focus on optimizing sampling noise schedules and their hyperparameters~\citep{lin2024common,xue2024accelerating,sabour2024align}, with most relying on heuristics or grid search for their selection.
Even without rigorously formulating the minimization of discretization error into an optimization problem, analyzing the relationship between local truncation error and noise levels can provide valuable insights to guide the tuning of sampling schedules (see Appendix D.1 in~\citet{karras2022elucidating} for more details).
An approach to circumvent the task of optimizing the noise schedules is to introduce an ensemble of expert denoisers, with each expert denoiser is responsible for a particular range of noise levels~\citep{balaji2022ediff,feng2023ernie}.
This way, the model capacity can be increased without increasing the computational complexity during the sampling process.

\subsection{Integration Scheme (Sampler)} The choice of the integration scheme also significantly impacts both sampling quality and efficiency in solving ODEs.
After simplifying the EDM, the resulting ODE takes a form similar to that of DDIM. 
In this context, even a straightforward application of Euler's method can achieve quality comparable to the original DDPMs within 100 steps, resulting in at least a 10-fold speed increase~\citep{song2021denoising}.
To further accelerate sampling while maintaining generation quality, several advanced integration schemes have been proposed.
These include restart sampling~\citep{xu2023restart}, Heun's method~\citep{karras2022elucidating}, higher order methods~\citep{song2020score,li2024accelerating}, linear multistep methods~\citep{liu2022pseudo}, exponential integrator (EI) based methods~\citep{zhang2023fast,lu2022dpm,lu2022dpmpp,zhao2024unipc}, and truncated Taylor methods~\citep{dockhorn2022genie}.
Built upon Eq.~\eqref{eq:simedmode}, Heun's method has been empirically found to achieve a balance between low discretization error and a small number of NFEs\citep{karras2022elucidating}.

Among these advanced integration schemes, DPM-solver~\citep{lu2022dpm,lu2022dpmpp} has emerged as a particularly influential approach, gaining widespread adoptions in various open-source frameworks, including latent diffusion~\citep{rombach2022stable} and Diffusers~\citep{von2022diffusers}. 
Specifically, DPM-Solver exploits the semi-linear structure of diffusion ODEs, which consist of a linear function and a nonlinear function of the data $\boldsymbol{x}$ (typically a neural network term).
By directly approximating the exact solutions of these ODEs, which involve an exponentially weighted integral of the denoising or noise prediction model, DPM-Solver generates high-quality samples with 10 to 20 NFEs.
Inspired by numerical methods for exponential integrators~\citep{hochbruck2010exponential}, DPM-Solver also incorporates higher-order terms to approximate the exponentially weighted integral, providing theoretical convergence guarantees.
Notably, Eq.~\eqref{eq:edmode} also exhibits a semi-linear structure, and with the change-of-variable trick on $\lambda=-\log(\sigma)$, DPM-Solver++~\citep{lu2022dpmpp} can be naturally applied within the EDM context.
Furthermore, DPM-Solver++ employs multistep methods that reuse previous values to approximate higher-order derivatives, which are empirically more efficient than single-step methods.

Beyond efficient samplers, distillation-based models~\citep{song2023consistency,salimans2022progressive,yin2024one} have gained popularity for their ability to achieve high-quality results in as few as five steps or even one step, albeit at the cost of additional student model training. 
While there are various techniques and variants of diffusion distillation, a comprehensive discussion of these methods is beyond the scope of this paper.


\section{Sampling Guidance}
\label{sec:sg}
During sampling, additional guidance techniques can be applied to direct the model towards a desired conditional distribution, thereby improving generation performance.
These techniques primarily fall into two categories: \textit{classifier guidance} and \textit{classifier-free guidance (CFG)}. 
Classifier guidance utilizes post-hoc noise-conditioned classifiers on pretrained unconditional models~\citep{dhariwal2021diffusion,song2020score}, while CFG involves interpolation or extrapolation between conditional and unconditional models~\citep{ho2022classifier}. 
Recent research has also explored combining both guidance methods~\citep{luo2024diff}.

In classifier guidance, the guided sampling process modifies the unconditional score $\nabla_{\mathbf{x}} \log q(\mathbf{x}(t))$ by incorporating additional classifier gradients through Bayes' rule:
\begin{equation}
\nabla_{\mathbf{x}} \log q(\mathbf{x}(t)|\mathbf{y})=\nabla_{\mathbf{x}} \log q(\mathbf{x}(t))+s\cdot\nabla_{\mathbf{x}} \log q(\mathbf{y}|\mathbf{x}(t)),
\label{eq:classifier}
\end{equation}
where $s$ is a scalar factor that controls the strength of the guidance, and $q(\mathbf{y}|\mathbf{x}(t))$ represents an auxiliary classifier that predicts the condition $\mathbf{y}$ given noisy inputs $\mathbf{x}(t)$.
The gradients with respect to $\mathbf{x}(t)$ are computed via back propagation.

Given that the sampling process can be viewed as ODE solution trajectories, as discussed in Sec.\ref{sec:sampling}. 
This perspective allows for a geometric interpretation of guidance techniques\citep{dieleman2023geometry}, offering insights into how they shape the model's sampling path.
In this geometric view, both the score and classifier gradients act as directional guides at each step of the ODE sampling process.
The classifier's influence effectively alters the sampling trajectory, steering it towards the desired conditional outcome. 
The scalar $s$ controls the influence of the conditioning signal on the sampling procedure by scaling the classifier gradient's contribution.

This classifier guidance approach extends beyond traditional classifiers, as the guidance term $q(\mathbf{y}|\mathbf{x}(t))$ can generalize to various forms of supervision. 
The ``classifier" can process categorical labels as in traditional classification, but also continuous-valued variables through contrastive models~\citep{nichol2021glide} or discriminators~\citep{kim23refining}.
This flexibility enables diverse applications, including inverse problems~\citep{daras2024survey}, where the supervision signal (guidance) comes from partial or degraded measurements of the target. 
While this generalized supervised guidance offers theoretical advantages, training reliable guidance models remains challenging in practice~\citep{song2022solving,levy2023controllable,chung2023diffusion,chung2022improving,song2023loss}.


Unlike classifier guidance which requires training separate classifiers that can overfit to noise patterns and dataset biases, CFG offers a simple yet effective alternative technique that avoids these limitations.
Building upon Eq.~\eqref{eq:classifier} and applying Bayes' rule to the classifier term $q(\mathbf{y}|\mathbf{x}(t))$, we obtain:
\begin{equation}
\nabla_{\mathbf{x}} \log q(\mathbf{x}(t)|\mathbf{y})=(1-s)\cdot\nabla_{\mathbf{x}} \log q(\mathbf{x}(t))+s\cdot\nabla_{\mathbf{x}} \log q(\mathbf{x}(t)|\mathbf{y}).
\label{eq:cfg}
\end{equation}
This formulation only involves data likelihood but not the classifier $q(\mathbf{y}|\mathbf{x}(t))$. 
$s$ is typically set greater than 1.
CFG can be further extended through composable guidance approaches~\citep{liu2022compositional}, which enable multiple conditioning signals to be combined during generation. 

Geometrically, CFG creates a trajectory that emphasizes the difference between conditional and unconditional paths. 
When $s>1$, the trajectory extrapolates beyond the conditional prediction in the direction opposite to the unconditional prediction, effectively concentrating samples on the conditional distribution while pushing them away from the unconditional one. 
This geometric perspective illuminates how CFG manipulates the vector field to enhance conditional features during sampling.

However, applying CFG throughout all noise levels is not necessary and can lead to issues:
At low noise levels, generation is sensitive to CFG~\citep{saharia2022photorealistic,lin2024common}, causing an over-exposure issue.
At high noise levels, it can misdirect ODE trajectories away from the data distribution, leading to a phenomenon called ``catastrophic mode drop''~\citep{kynkaanniemi2024applying}.
To address these challenges, several approaches have been proposed.
In Imagen~\citep{saharia2022photorealistic}, a cosine scheduled CFG coefficient and dynamic range are applied to overcome the over-exposure issue. 
Similarly,~\citet{lin2024common} proposed a rescaling technique at each CFG step to prevent pixel overexposure caused by large guidance scaling factors.
More recently,~\citet{karras2024guiding} propose to replace the unconditional model with a weaker conditional model that tends to model a more spread-out representation of the entire training dataset. 
The strong and weak conditional models exhibit different degrees of uncertainty towards similar regions. 
Areas where they disagree most are likely under-fit regions. 
This difference can correct the sampling trajectory direction, pushing it towards better samples. 
This approach shares similarities with contrastive decoding in language models~\citep{li2023contrastive}, which leverages the differences between a stronger (higher-quality) model and a weaker (lower-quality) model to enhance generation quality.

Beyond training-free guided sampling for post-hoc editing or controllable generation, alternative paradigms exist, such as conditional adaptation~\citep{zhang2023adding,wu2024music} and latent space searching~\citep{novack2024ditto}. However, these approaches are less relevant to the solution of the sampling ODE and fall outside the scope of this review.

\section{Architecture Design}
\label{sec:ad}
\subsection{Data Representation}
In the audio domain, diffusion models can operate on various input feature representations, each with its own advantages and trade-offs. 
A vanilla choice is raw waveforms~\citep{rouard2021crash,pascual2023full,kong2020diffwave,san2024discrete,gao2023e3}. 
However, raw waveforms often exhibit temporal redundancies, leading to long sequence lengths for full-band audio.

Spectral features, such as wavelet or constant Q-transforms, offer improved scalability in time length while preserving the harmonic structure of audio signals. 
These representations are generally easier for the diffusion process to model~\citep{huang2023noise2music,moliner2023solving,lee2021nuwave,hoogeboom2023simple,hawthorne2022multi}.
Another effective approach is the complex spectrogram with exponential amplitude transformation ($\tilde{c} = \beta|c|^\alpha e^{i\angle(c)}$), which both preserves phase information critical for high-fidelity audio reconstruction and addresses the heavy-tailed distribution of STFT amplitudes.
It was first applied in speech processing~\citep{welker2022speech}, and was later shown to be effective in sound effects~\citep{zhu2023edmsound} and music generation~\citep{nistal2024diff}.

Similar to developments in the image domain, discrete or continuous VAE latent representations have also gained popularity in audio diffusion models~\citep{liu2023audioldm,huang2023make,ghosal2023tango,yang2023diffsound,evans2024fast,schneider2023mo}.
These VAEs, often built upon perception-inspired GAN architectures~\citep{kong2020hifi,defossez2023high,zeghidour2021soundstream}, provide semantically meaningful and compressed representations of audio data. 
By extracting these more compact feature maps, VAEs effectively decouple perceptual information from the generative modeling process, allowing diffusion models to focus primarily on semantic content. 
This approach reduces computational costs and improves efficiency in both optimization and inference~\citep{rombach2022stable}.

\subsection{Neural Networks}  
There have been notable developments in neural network backbones for audio diffusion models. 
DiffWave~\citep{kong2020diffwave} adopts a bidirectional dilated convolutional neural network from WaveNet~\citep{aaron2016wavenet}. 
This architecture was chosen for its large receptive fields, which are particularly beneficial in processing waveform-based audio signals. 
Convolutional U-Nets~\citep{ronneberger2015u} and their variants~\citep{lin2017refinenet} have become the preferred neural network architectures for diffusion models due to their multi-resolution analysis capabilities.
These networks are good at capturing both global and local context information, while preserving fine-grained details through skip connections. 
This effectiveness has been demonstrated in numerous studies across both signal and latent spaces from images to audio~\citep{ho2020denoising,nichol2021improved,dhariwal2021diffusion,karras2022elucidating,de2021diffusion,yang2023diffusion,kingma2021variational,saharia2022photorealistic,rombach2022stable,liu2023audioldm,welker2022speech,richter2023speech}.
The U-Net model uses an encoder-decoder structure with two main parts: a downsampling path and an upsampling path, with skip connections between these paths preserving low-level features and leading to more accurate reconstructions. 

Diffusion transformers (DiTs)\citep{peebles2023scalable,bao2022all,bao2023one,hoogeboom2023simple,zheng2024fast,chen2024pixartalpha,gao2023masked}, on the other hand, utilize the remarkable scaling properties of transformers, which have demonstrated significant performance gains with increased model size, computational resources, and data volume in both language~\citep{kaplan2020scaling} and vision~\citep{dosovitskiy2021an} domains.
DiTs follow a design methodology similar to that of vision transformers (ViT). 
It begins with a ``patchify" step, where the input noisy data (e.g., images, spectrograms, or VAE latents) are converted into a sequence of tokens and combined with standard ViT positional encodings. 
This token sequence is then processed through a series of transformer blocks. 
In the final step, the processed tokens are decoded back into the original spatial layout to produce the diffusion output.

While U-Nets and transformers are prominent choices for diffusion models, alternative architectures have emerged to combine their strengths and address their limitations. 
The hourglass transformer~\citep{nawrot2022hierarchical}, proposed for pixel space diffusion models~\citep{crowson2024scalable}, builds upon the multi-resolution inductive bias inherent in U-Nets while leveraging the scalability of transformers. 
Additionally, state space model (SSM) based backbones~\citep{gu2023mamba,gu2022efficiently}, an efficient alternative for general-purpose neural sequence modeling, have been introduced as attention-free diffusion architectures~\citep{yan2024diffusion,teng2024dim}. 
These SSM-based models address the quadratic computational requirements that arise with long sequences in traditional transformer architectures.

\subsection{Conditioning} 
\label{sec:cond}
Many audio applications can be effectively framed as conditional or controllable generation tasks, albeit with varying input dimensions. 
For instance, text-to-audio generation involves producing spectrograms or waveforms based on given sequences of text embeddings, while speech enhancement aims to recover clean speech waveforms conditioned on their noisy versions. 
Within the framework of score modeling, these tasks can be represented using a conditional score model $\nabla \log p(\mathbf{x}|\mathbf{y})$, where $\mathbf{x}$ denotes the target audio and $\mathbf{y}$ the conditioning information.
To leverage the scalability and high-quality generation capabilities of diffusion models across this diverse range of audio tasks, it is crucial to incorporate appropriate conditioning techniques into the neural networks. 

\paragraph{In-Context}
One straightforward approach is through sequence concatenation, which appends conditional embeddings as additional conditional tokens to the input sequence, treating them identically to data tokens~\citep{bao2022all,peebles2023scalable}. 
After processing through the final block, the data tokens are retained while the conditioning tokens are removed from the sequence. 
A more advanced approach~\citep{bao2023one} integrates representations of different modalities into a single sequence of tokens and apply different noise levels to different modalities.
In this method, the score model explicitly fits all relevant distributions within one model. 
This unified approach enables both individual modality sampling and joint sampling at inference time without requiring additional training.
However, this method has limitations including increased computational cost with longer sequences and difficulty in controlling the strength of the conditioning influence on generation.

Another common technique is channel concatenation, where conditional information is incorporated by concatenating it along the feature dimension while aligning in the time dimension.
For audio-based applications such as text-to-speech synthesis, this strategy enables precise temporal control through the use of phoneme sequences and masked spectrograms~\citep{le2024voicebox,popov2022diffusionbased}. 
Such temporal alignment is crucial for tasks requiring fine-grained synchronization between conditional information and audio output.

\begin{figure}
    \centering
    \includegraphics[width=0.99\columnwidth]{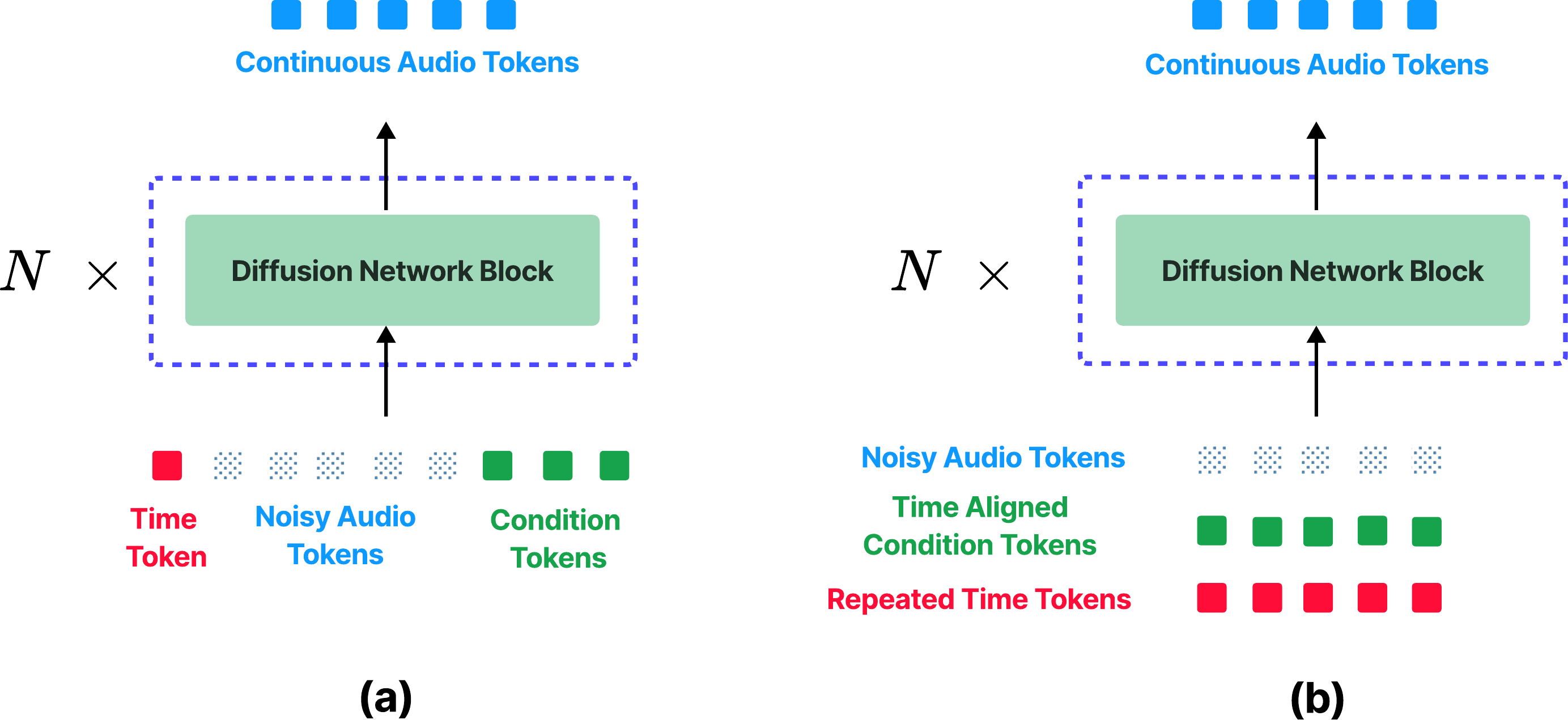}
    \caption{In-context conditioning: (a) sequence concatenation, 
    where time tokens are embeddings output from MLP layers;
    (b) time aligned channel concatenation. In this example, we use a denoising objective; in flow matching, the output can be the ``velocity''.}
    \label{fig:incontext}
\end{figure}


\paragraph{Modulation} Another common approach for incorporating conditional information into diffusion models involves adaptive layer normalization (AdaLN) or feature-wise linear modulation (FiLM) layers~\citep{perez2018film}. 
These techniques, widely adopted in GAN generators~\citep{brock2018large,karras2019style}, modulate the model's behavior based on input conditions. 
Additionally, gating mechanisms~\citep{van2016conditional,aaron2016wavenet}, inspired by LSTM multiplicative units~\citep{hochreiter1997long}, offer another approach for conditional control in diffusion models.
These modulation-based methods allow for the integration of various types of conditional information.
Noise level $\sigma$, introduced in Sec.~\ref{sec:trainnoise}, is typically combined with additive positional encoding and then fed into a learnable MLP to generate noise embeddings. 
Other conditional inputs can include timing conditions\citep{evans2024fast}, one-hot class labels~\citep{pascual2023full}, contrastive language-audio embeddings~\citep{wu2023clap,liu2023audioldm}, or combinations of such vectors~\citep{xu2024vasa}.
In AdaLN, the sum of these conditional vectors and noise embeddings is processed by a shallow MLP to learn dimension-wise scaling and shifting parameters. 
Gating mechanisms, on the other hand, employ linear layers and gated activation units. 
These parameters are used to modulate features within the score network, allowing the model to adapt its behavior based on the input conditions. 

\begin{figure}[!t]
\centering
\includegraphics[width=0.77\columnwidth]{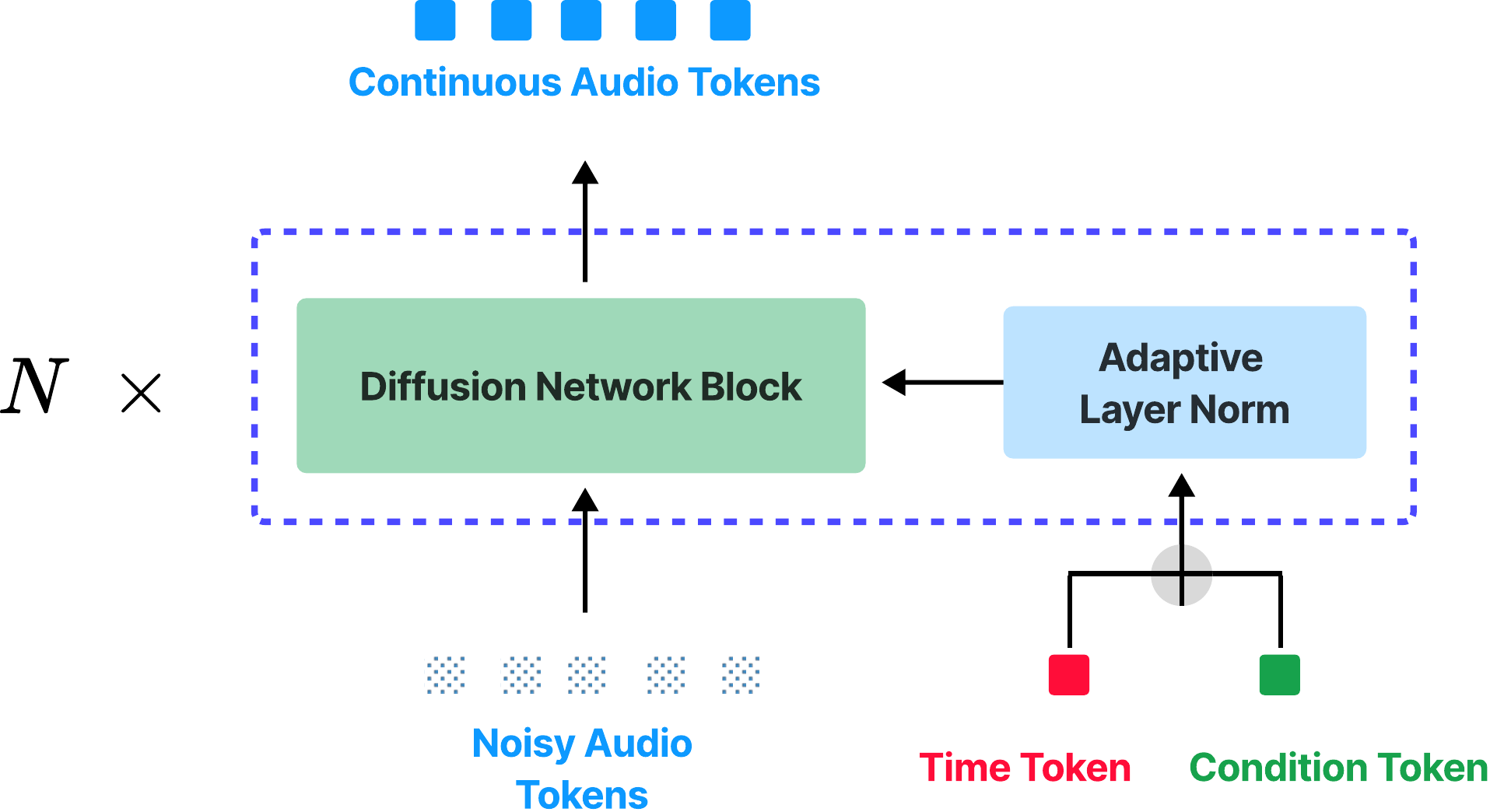}
\caption{Adaptive layer norm modulation.
}
\label{fig:modulation}
\end{figure}

\paragraph{Cross Attention} Within each U-Net or DiT block, cross-attention layers serve as a key conditioning mechanism to enable diffusion models to incorporate external information. This mechanism establishes direct connections between the model's internal representations and conditional inputs, making it particularly valuable for integrating multi-modal sequential conditions, such as text tokens of varying lengths.
In implementation, cross-attention layers are typically placed after self-attention components within transformer blocks, following the design principles established by~citet{vaswani2017attention}.
This conditioning strategy has proven effective across different domains, including image generation\citep{saharia2022photorealistic,rombach2022stable} and audio synthesis~\citep{lovelace2023simple,kreuk2022audiogen,evans2024stable}.

\begin{figure}[!t]
\centering
\includegraphics[width=0.77\columnwidth]{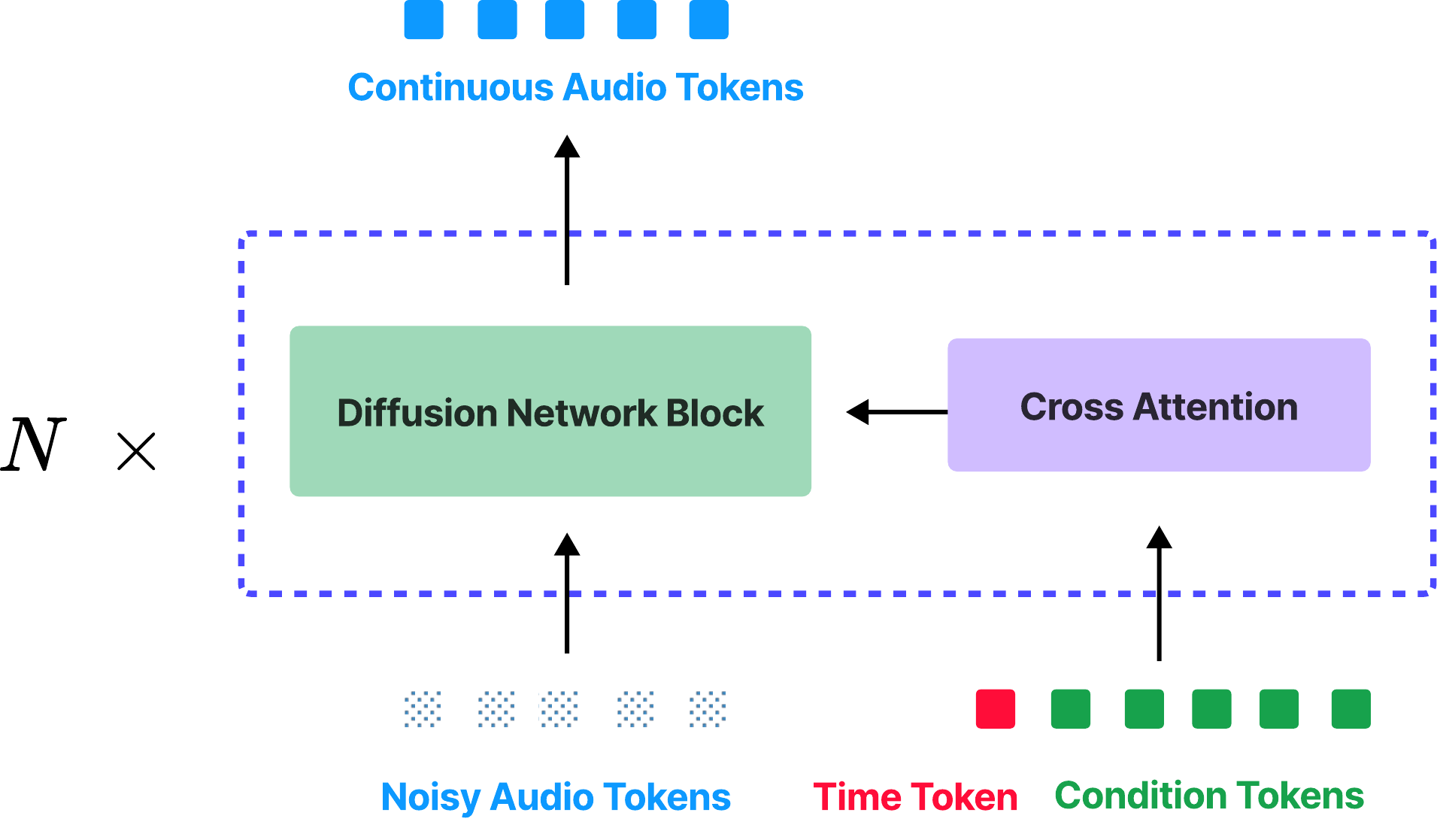}
\caption{Cross attention conditioning.
}
\label{fig:modulation}
\end{figure}

\paragraph{Boundary Condition}
In traditional diffusion modeling, we work with fixed noise and data distributions, typically starting from Gaussian noise as the initial distribution and use the data distribution as the target distribution, while incorporating external conditioning information through neural networks.
However, an alternative approach is to modify either the initial or target distribution itself, which can simplify generative modeling across various scenarios.
We term this strategy ``boundary condition" in this survey.
This strategy offers a powerful method to guide the diffusion process towards more plausible outputs, potentially reducing the number of required sampling steps and enhancing the overall quality of the generated content.
One idea is to change the target data as the residuals between the clean data and an initial rough estimation from a deterministic predictor~\citep{whang2022deblurring}.
In this prediction-then-generation strategy, the reverse diffusion process is simplified and essentially becomes a refinement module leveraging the the initial estimation. 
Another idea is to set the mean and standard deviation parameters of the initial normal distribution to a more informed estimate from the target data.
This change requires a modification to the perturbation kernel defined in Eq.~\eqref{eq:noising}, which has a closed-form solution given drift coefficient, diffusion coefficient, and initial distribution~\citep{sarkka2019applied}.
More aggressively, the Gaussian perturbation process in diffusion models can be generalized with other kernels~\citep{yen2023cold,hoogeboom23blurring,rissanen23generative,xu2022poisson}; This, however, is beyond the scope of this review.

The aforementioned conditioning techniques can also be combined~\citep{xu2024vasa}.
However, these methods vary significantly in their computational complexity~\citep{peebles2023scalable}.
The computational overhead ranges from minimal to substantial across different approaches. 
Channel concatenation and changing the distribution initial conditions are the most efficient, 
as they only increase computational cost linearly with respect to the number of additional input channels and only affect the initial convolutional layer.
Sequence concatenation maintains reasonable efficiency, with additional FLOPs determined by the product of conditioning sequence length and model hidden dimension~\citep{peebles2023scalable}. 
The FiLM layer method incurs moderate computational costs through its parameter-generating MLP, resulting in a modest increase in total FLOPs~\citep{peebles2023scalable}.
Cross-attention using transformer architectures represents the most computationally demanding approach, as its multi-head attention operations significantly increase FLOPs, particularly when processing longer sequences~\citep{peebles2023scalable}.

\chapter{Application}

\label{sec:app}
In this section, we review the application of diffusion models to three main audio tasks, categorized based on the strength of alignment between the given conditions and target generation samples. 
We discuss, in order of increasing conditional strength: (1) Text-to-audio (TTA), specifically general sound or music generation; (2) Text-to-speech synthesis (TTS); and (3) Source separation and speech restoration.

TTA and TTS are both tasks that generate audio content from textual input, yet they exhibit distinct properties and utilize different levels of textual information. 
TTA uses semantic concepts defined in the text to generate sound events similar to text-to-image generation, and it lacks a direct and exact correspondence between text and audio. In contrast, TTS maps phonemes parsed from the text to acoustic features and waveforms with a much more direct mapping between text and speech.
Applications in source separation and speech restoration represent the strongest conditional alignment, operating at the time-frequency bin level or waveform sample level given corrupted audio input.
Our discussion emphasizes the real challenges and the practical aspects of architecture design in these specific tasks.

\section{Text-to-Audio}


Text-conditioned audio generation is an emerging field that synthesizes sound effects or music directly from text descriptions. 
It offers a novel alternative to traditional methods of searching sounds from audio libraries.
Early work on label-based sound effects generation based on diffusion models has achieved high-fidelity performance. 
Rouard and Hadjeres~\citep{rouard2021crash} proposed a score-based generative model for drum sound generation, capable of producing full-band waveforms of half a second long. 
Their approach utilizes a U-Net architecture with FiLM layers for label conditioning. 
Similarly,~\citet{pascual2023full} introduced a score-based model that operates on waveforms too, generating full-band audio up to 10 seconds in length from general sound event labels. 
\citet{zhu2023edmsound} introduced EDMSound, another label-based end-to-end score-based sound generator operating on complex spectrograms. 
EDMSound treats the real and imaginary components as separate channels in a 2-channel representation, enabling modeling of both magnitude and phase information.

Extending label-based generation to arbitrary textual descriptions presents two key challenges in the field: (1) Modeling the inherent complexity of long continuous waveform data to produce coherent and diverse audio that faithfully represents textual inputs, particularly in music generation; (2) Addressing the limited availability of large-scale, high-quality text-audio paired datasets in the public domain, especially for sound effects generation. 
This scarcity significantly influences text-audio alignment and fidelity of TTA systems.

To address the challenge on long-form audio generation, earlier work focused on auto-regressive generation methods, such as SpecVQGAN in the spectrogram domain~\citep{iashin2021taming,liu2021conditional} and AudioGen~\citep{kreuk2022audiogen} using encodec based acoustic tokens~\citep{defossez2023high}. 
More recently, high-quality LDMs have enabled the production of longer audio samples in a non-autoregressive fashion~\citep{yang2023diffsound, ghosal2023tango, huang2023make, liu2023audioldm}. 
These methods typically employ a VAE to compress the waveform or spectrogram into a more compact representation, followed by text-conditioned diffusion models operating in the latent domain.
\citet{yang2023diffsound} first introduced DiffSound, an LDM-based text-to-audio system utilizing discrete diffusion models. 
Their approach employs audio tokens derived from a spectrogram VQ-VAE~\citep{iashin2021taming}. 

While discrete diffusion models existed in early stages of development, continuous VAE-based autoencoders are more commonly used in current research~\citep{ghosal2023tango, huang2023make, liu2023audioldm,liu2024audioldm}.
\citet{huang2023noise2music} introduced Noise2Music, which employs two cascaded systems operating on waveforms and spectrograms. 
This approach, consisting of a base diffusion model and a cascader (upsampler for waveform model and vocoder for spectrogram model), progressively generates 30-second waveforms sampled at 24 kHz from text prompts.
\citet{schneider2023mo} introduced Moûsai, a system employing a diffusion-based autoencoder~\citep{preechakul2022diffusion} for audio generation. 
This approach compresses spectrogram magnitudes, reducing waveform dimensions by a factor of 64. Notably, Moûsai achieves minute-long, full-band stereo music generation.
\citet{li2024jen} introduced Jen-1, which is based on a continuous VAE for stereo music sampled at 48 kHz with a latent rate of 125. 
Jen-1 incorporates various masking strategies in its U-Net block, enabling inpainting and continuation capabilities alongside text-based music generation.
The LDM-based Stable Audio series \citep{evans2024fast, evans2024stable, evans2024long} demonstrates significant advancements in audio generation. 
These models can produce stereo audio at a 44.1 kHz sampling rate and excel in generating long-context music pieces of up to approximately 5 minutes based on text prompts, supporting variable-length generation through timing conditions.
\citet{evans2024fast, evans2024stable, evans2024long} trained Stable Audio on large-scale text-music datasets with improved VAE and diffusion backbones, achieving state-of-the-art performance in both musical quality and text-audio alignment among open text-based music generation models.

To tackle the challenge of limited high-quality paired text-audio datasets, one effective method involves cross-modal embedding alignment through contrastive language-audio pretrained embeddings~\citep{wu2023clap,elizalde2023clap}.
This approach leverages representations from contrastive learning where audio and text embeddings are aligned in a shared space.
AudioLDM~\citep{liu2023audioldm} implemented this by conditioning diffusion models on CLAP audio embeddings during training, while enabling text-guided generation at inference time through the aligned text embeddings.
Similarly, CLIPSonic~\citep{dong2023clipsonic} leveraged CLIP~\citep{radford2021learning} to train on image embeddings from unlabeled videos and to enable text-to-audio generation through the aligned text encoder, overcoming the paired data requirement.

Another approach involves transfer learning from robust pre-trained vision models. 
Given that text-to-image LDMs are trained on 400M image-text pairs~\citep{rombach2022stable}, 
models like Auffusion~\citep{xue2024auffusion} and Riffusion~\citep{forsgren2022riffusion} leverage their strong generation capability and cross-modal alignment by finetuning these models on text and spectrogram pairs. 
This approach effectively transfers the pre-trained knowledge to text-to-audio and text-to-music generation while requiring significantly less training data and computational resources compared to training from scratch.
Data augmentation with synthetic data is another widely adopted strategy.
\citet{huang2023make} introduced Make-An-Audio, which addresses text-audio alignment challenges through a prompt enhancement approach. 
Their method employs an audio captioner to generate textual descriptions for audio samples and an audio-text retriever to identify well-aligned text-audio pairs. 
These pairs are then refined through a reprogramming procedure that creates diverse compositional prompts by sampling and combining sound events according to templates. 
The resulting text embeddings are integrated into the diffusion model via cross-attention layers in the U-Net backbone.
\citet{yuan2024improving} proposed Sound-VECaps, which further integrates vision captioners to generate detailed captions for AudioSet videos~\citep{gemmeke2017audio}. 
These captions are then refined using Large Language Models (LLMs), significantly improving performance on complex prompts.

Additional common data augmentation techniques include superimposing existing audio pairs and concatenating their captions~\citep{liu2023audioldm,ghosal2023tango,kreuk2022audiogen}. 
Tango~\citep{ghosal2023tango} refined this approach by introducing a mixup-style data augmentation method that combines two audio samples based on their sound pressure levels. 
This technique, inspired by~\citet{tokozume2018learning}, creates new training examples by blending audio pairs with appropriate weights to ensure a balanced representation of both source signals.
Building upon Retrieval-Augmented Generation (RAG)~\citep{blattmann2022retrieval},~\citet{yuan2024retrieval} enhanced AudioLDM by incorporating a retrieval mechanism that identifies relevant text-audio pairs from a database, extracts their embeddings, and uses these as additional conditioning signals during the diffusion generation process.

An additional challenge in music generation is generating rich musical structures conditioned on musical attribute inputs.
Several approaches have been proposed to address this challenge. 
\citet{hawthorne2022multi} introduced a multi-instrument spectrogram diffusion model based on spectrogram domain DiT, offering note-level control over both composition and instrumentation.
Their model supports arbitrary-length music generation by employing a second encoder stack that processes previously generated segments, ensuring smooth transitions between consecutive sections.
Recent advances in music generation systems have incorporated techniques from latent diffusion models with text conditioning.
Mustango~\citep{melechovsky2024mustango} developed a music-domain knowledge-informed U-Net guidance module that incorporates music-specific conditions such as beats and chords, alongside textual descriptions, through cross-attention layers. 
MusicLDM~\citep{chen2024musicldm} improved both quality and novelty of generated music, as well as text-music correspondence, by utilizing beat-synchronous data and mix-up augmentation techniques. 
Diff-a-Riff~\citep{nistal2024diff} extends input data beyond text prompts by generating instrumental accompaniments conditioned on musical inputs.  
The JEN-1 Composer~\citep{yao2023jen} combines LDM with progressive curriculum training to enable track-by-track music generation, allowing users to iteratively produce and refine individual instrumental parts while maintaining coherence across the complete composition.


\section{Text-to-Speech}

Neural network-based TTS systems~\citep{tan2021survey} aim to synthesize natural speech from text input with a neural network model. 
Contemporary TTS architectures typically comprise three fundamental components: (1) a text analysis module (or text encoder) that converts input text into linguistic features such as phoneme embeddings, (2) an acoustic model (or decoder) that generates acoustic features like mel-spectrograms from these linguistic representations, and (3) a vocoder that synthesizes the final waveform from the acoustic features.
Phoneme-level text encoders are often used as common choices for the first component. 
These encoders process phoneme sequences through sequence modeling, utilizing LSTM or transformer-based layers. 
The encoded phoneme embedding sequences are then expanded by a duration-informed length regulator~\citep{ren2019fastspeech}. 
Previous diffusion-based TTS systems have primarily focused on developing acoustic models~\citep{popov2021grad,huang2022fastdiff,guan2024reflow} and vocoders~\citep{chen2021wavegrad,kong2020diffwave} for non-autoregressive generation. 

\subsection{Acoustic Models}
The acoustic model serves as a crucial component, bridging the gap between textual and audio modalities.
Recent advancements in diffusion-based and flow-based models have significantly improved this component's performance.
In Diff-TTS~\citep{jeong2021diff}, expanded phoneme embeddings are integrated through a gating modulation method to incorporate sequence-level phoneme controls into the training of the DDPMs.
Building on the success of normalizing flow-based TTS systems~\citep{kim2020glow}, which utilize expanded phoneme embeddings as latent priors and a monotonic alignment-based duration predictor, Grad-TTS replaces the Glow generator~\citep{kingma2018glow} with score-based models to generate mel-spectrograms, achieving superior performance.
DiffVoice~\citep{liu2023diffvoice} introduces an acoustic model based on LDMs, incorporating channel-concatenated phoneme embeddings and repeated time embeddings alongside the noised VAE latent as input to the score model. 
Flow matching based acoustic models~\citep{kim2024p,guo2024voiceflow,guan2024reflow} have demonstrated high-quality synthesis within a limited number of steps.
In particular,~\citet{kim2024p} proposed P-Flow, which features a mask-based approach for speaker adaptation and a flow matching-based decoder. 
This decoder concatenates noisy data with phoneme embeddings along the sequence dimension, outperforming traditional speaker embedding conditioning.
To integrate more control, SimpleSpeech~\citep{yang2024simplespeech} adopts a sequence concatenation approach, combining text-level embeddings, time tokens, speaker embeddings, and duration embeddings with scalar latents along the sequence dimension, similar to U-ViT~\citep{bao2022all}, and employs a GPT-2 style transformer~\citep{radford2019language} as diffusion backbone.
Guided-TTS~\citep{kim2022guided,choi2023prior} introduces an alternative approach, employing a guidance-based method for the acoustic model.
This system first trains a DDPM model on large-scale untranscribed data, then generates the mel-spectrogram guided by phoneme classifiers and speaker embeddings.

More recently, research in speech synthesis has shifted focus towards inference efficiency and controllable generation. 
The computational demands of diffusion models, particularly their Number of Function Evaluations (NFEs), have prompted efforts to enhance real-time feasibility. 
Various approaches have been explored to accelerate these models, including improved sampling processes~\citep{chen2023lightgrad,huang2023fastdiff,lam2022bddm}, conversion of reverse diffusion into GANs~\citep{xiao22tackling,liu2022diffgan,vovk2022fastgt}, and distillation-based methods~\citep{huang2022prodiff,ye2023comospeech}.
Acoustic models must also account for rich paralinguistic information such as speaker identity, prosody, and emotions. 
Integrating these attributes for fine-grained control has become another crucial aspect of acoustic modeling. 
For instance, EmoDiff~\citep{guo2023emodiff} employs soft emotion labels with classifier guidance upon GradTTS to directly control arbitrary emotions with intensity. Grad-stylespeech~\citep{kang2023grad} modulates expanded prior phoneme embeddings and UNet blocks through style vectors encoded from reference speech.
Notably, some approaches like Style-TTS~\citep{li2024styletts} and Prosody-TTS~\citep{huang2023prosody} utilize the diffusion process to model diverse speech prosody or styles rather than for acoustic modeling itself. 
This strategy significantly enhances generation efficiency compared to typical diffusion TTS models while still benefiting from diverse speech generation capabilities.

\subsection{Vocoders}
Vocoders typically exhibit stronger alignment between input (acoustic features) and output (waveforms) compared to acoustic modeling tasks. 
\citet{chen2021wavegrad}  introduced WaveGrad, a score-based vocoder conditioned on mel-spectrograms via FiLM layers.
Their comparison of discrete and continuous noise-level conditioned models revealed superior performance in the continuous variant, enabling more flexible noise scheduling during inference. 
Concurrently, \citet{kong2020diffwave} proposed DiffWave, a vocoder based on discrete-time diffusion models that employs gating mechanisms for mel-spectrogram conditioning.
Subsequently,~\citet{lee2022priorgrad} introduced PriorGrad~, which leverages a data-dependent prior derived from the normalized frame-level energy of mel-spectrograms. 
This approach extracts energy values from each frame of the mel-spectrogram and normalizes them to create a variance distribution that better reflects the natural acoustic patterns in speech. 
It provides the diffusion model with a more informative prior that aligns with the natural energy contours of audio signals, making the generation process more efficient and effective than using a standard Gaussian prior.
Building upon PriorGrad, \citet{koizumi2023specgrad} developed SpecGrad, refining the prior conditioning method by integrating signal power and spectral envelope techniques, thus further enhancing vocoder quality.

\subsection{End-to-End}
More recently, the speech synthesis paradigm has shifted to training on significantly larger datasets, scaling up from hundreds to hundreds of thousands of hours.
This paradigm shift has led to the adoption of various approaches, including token-based language modeling~\citep{zhang2024speechtokenizer,wang2023neural,lajszczak2024base,borsos2023soundstorm,borsos2023audiolm,betker2023better}, diffusion models~\citep{le2024voicebox,anastassiou2024seed,eskimez2024e2,liu2024generative}, or a combination of both~\citep{anastassiou2024seed}.
A key strategy for scaling up is to simplify speech modeling with minimal supervision~\citep{gao2023e3,kharitonov2023speak,le2024voicebox}.
VoiceBox~\citep{le2024voicebox} trains on a text-guided speech infilling task, generating masked speech given surrounding audio and text transcripts through channel concatenation of masked mel-spectrograms, phoneme-level embeddings, and noisy speech using conditional flow matching.
While contemporary approaches such as~\citep{shen2023naturalspeech} explicitly model fine-grained speech attributes, VoiceBox simplifies the process by eliminating explicit paralinguistic labels (e.g., speaker or emotion) and instead captures these characteristics implicitly through in-context learning from surrounding audio.
E2 TTS~\citep{eskimez2024e2} further simplifies this approach by removing phoneme modeling. 
Spear TTS~\citep{kharitonov2023speak} conceptualizes TTS as two sequence-to-sequence tasks: translating text to semantic tokens using language models, and then translating semantic tokens to acoustic tokens through generative models. 
In the E3 TTS,~\citet{gao2023e3} summarize that text normalization, input unit selection, and alignment modeling significantly increase the complexity of existing TTS systems. 
They propose relying solely on diffusion models for end-to-end speech generation by directly using pretrained BERT embeddings and conditioning them through cross-attention layers in the U-Net diffusion backbone. 
Similarly, sample-efficient speech diffusion~\citep{lovelace2023simple} applies ByT5~\citep{xue2022byt5} combined with positional embeddings and cross-attention into a latent transformer.

\section{Audio Restoration and Source Separation}
\subsection{Audio Restoration}
\label{sec:audiorestore}
Audio restoration involves transforming degraded audio signals into high-quality clean signals, such as speech enhancement and music restoration.  
In this section, we focus on the single-channel scenario.

Speech enhancement includes various sub-tasks such as dereverberation, de-clipping, and bandwidth extension~\citep{liu2024audiosr,yu2023conditioning}.
Recent advancements in speech enhancement have been largely driven by two categories of neural network-based approaches: \textit{predictive models} and \textit{generative models}, which have significantly improved audio quality~\citep{lemercier2024diffusion}.
Predictive models directly estimate clean speech from corrupted input by minimizing a point-wise loss function between the model output and the clean target signal ~\citep{luo2019conv,hu2020dccrn,park2022manner}.
However, this approach has shown limitations, including susceptibility to residual noise and artifacts~\citep{pirklbauer2023evaluation}, as well as difficulty generalizing to unfamiliar noise types and reverberation conditions~\citep{su2020hifi,gonzalez2023assessing}.

Generative models use a different learning paradigm. 
They aim to first explicityly or implicitly model a conditional or unconditional distribution of the clean audio data, and then sample from the learned distribution to achieve enhancement. 
Unlike predictive models, which produce deterministic outputs, generative approaches can sample multiple plausible solutions from a learned distribution, providing both diverse enhancement options and implicit uncertainty estimation through sample variance~\citep{lemercier2024diffusion}.
The generative approach is shown to be more favored in reference-free metrics of speech quality/naturalness, but is more penalized in reference-based metrics under low SNR conditions~\citep{de2023behavior}.
Among early generative approaches, GAN-based methods~\citep{pascual2017segan,su2020hifi} demonstrated strong performance, where a generator network is trained to output clean speech together with an adversarial discriminator that is trained to classify the generator output from clean speech. 
More recently, diffusion models have surpassed GANs in popularity and performance due to their ability to generate higher-quality samples while offering more stable training dynamics~\citep{zhang2021restoring,lu2021study,lu2022conditional}.

Diffusion-based generative models for audio restoration generally follow three main approaches. One straightforward method is to frame restoration as conditional generation, where the corrupted audio serves as the input condition. 
This approach leverages the conditioning strategies discussed in Sec.~\ref{sec:cond}. 
For instance,~\citet{zhang2021restoring} proposed ModDW, a DiffWave-based method that employs noisy mel-spectrograms and a CNN-based upsampling conditioner to generate clean waveforms.
\citet{serra2022universal} introduced UNIVERSE, a waveform domain score-based model with auxiliary mixture density network (MDN) losses and conditioner design, capable of producing high-quality enhanced speech across various audio degradations. 
Building upon this,~\citet{Scheibler2024universepp} developed UNIVERSE++, which incorporates architectural improvements and applies additional adversarial losses on the conditional decoder during training.
UNIVERSE++ also implements a low-rank adaptation scheme~\citep{hu2022lora} with a phoneme fidelity loss at the fine-tuning stage to improve content preservation.
Moreover, as this approach involves conditional generation, additional information from the noisy speech can be leveraged to improve performance. 
\citet{hu2023noise} propose a method that utilizes a trained noise classifier to extract noise embeddings, which are then used as conditioning inputs for diffusion-based speech enhancement models. 
This approach enables more targeted and effective enhancement by incorporating specific noise characteristics into the generation process.

In diffusion-based speech enhancement models, an alternative approach involves substituting the mean of the initial distribution with the noisy speech. 
Consequently, the mean of the degraded signal throughout the diffusion process becomes a linear interpolation between clean and noisy speech, with progressively increasing Gaussian noise. 
This technique is also referred to as interpolation diffusion models (IDM)~\citep{guo2023vp}.
\citet{lu2021study} initially proposed DiffuSE, followed by its extension CDiffuSE~\citep{lu2022conditional}. 
These models adapt the DDPM-based DiffWave architecture to incorporate the aforementioned diffusion setup.
Score-based generative models for speech enhancement (SGMSE)~\citep{welker2022speech,richter2023speech,lemercier2023analysing}, operate on complex spectrograms, utilize an Ornstein-Uhlenbeck process~\citep{karatzas1991brownian} in the SDE framework. 
\gz{In this approach, the drift term guides the mean of the Gaussian process from clean speech towards noisy speech.}
\citet{lay2024single} propose a two-stage approach for SGMSE. 
They first train a SGMSE module using the denoising score matching objective, then fine-tune it using a regression loss between its estimate and the clean speech target.
This fine-tuning method specifically corrects both discretization errors (from step discretization) and prediction errors (from the score model) that occur when using a numerical solver such as Euler-Maruyama. 
By backpropagating through only the last reverse diffusion step, they achieve a significant reduction in required sampling steps while maintaining computational efficiency.
\citet{gonzalez2024diffusion} conducted a comprehensive investigation into the design of score-based frameworks, focusing on noise schedulers and samplers. 
Their findings demonstrated that these models outperform state-of-the-art predictive approaches in both matched and mismatched conditions, with particularly notable results when trained on diverse databases. 
In following work,~\citet{gonzalez2023investigating} adapted the SDE to be compatible with the EDM framework. 
They further examined the impact of the drift term by introducing an additional shift coefficient in preconditioning. 
Their results suggest that the contribution of the drift term to improving speech enhancement performance is not significant.
Further investigations have explored various design choices, including diverse drift terms~\citep{lay2023reducing, gonzalez2023investigating}, varying SDE variance scales~\citep{lay2024analysis}, and alternative diffusion processes~\citep{richter2024investigating,jukic2024schr,yen2023cold,saito2023unsupervised}. 

The third approach, termed ``prediction-then-generation'', leverages advancements from both predictive and generative models~\citep{lemercier2023storm, wang2023cross, sawata2023diffiner}.
This hybrid method has the potential for more robust and higher-quality speech enhancement. 
In this framework, predictive models initially estimate clean speech, after which diffusion models refine the output to enhance naturalness, as discussed in Sec.~\ref{sec:cond}.
This two-stage process combines the strengths of both model types, potentially leading to superior results. 
SRTNet~\citep{qiu2023srtnet} introduces this two-stage approach: noisy speech is processed by a predictive model, leaving the residual component as the training target for subsequent diffusion models. 
This strategy leverages the observation that modeling the residual between the initial prediction and the clean speech is less complex than modeling the entire enhancement process from scratch. 
Consequently, SRTNet demonstrates faster convergence compared to models that enhance noisy speech from scratch.
In Storm~\citep{lemercier2023storm}, the initial Gaussian distribution's mean is set as an estimate from predictive models. 
Building upon this concept,~\citet{tai2023revisiting} proposed an extension that incorporates non-parameterized ``classifier" guidance at sampling time. 
This guidance equivalently interpolates the estimated intermediate output from DDPM.
Furthermore, they reuse the predictive model for additional refinement. 
In their subsequent work on diffusion-dropout speech enhancement (DOSE),~\citet{tai2024dose} propose using an adaptive prior derived from the noisy speech as the initial distribution: either a clean speech estimation from the noisy input or a combination of this estimation with the original noisy speech.
DOSE also introduces a novel training strategy that incorporates conditional information more effectively by selectively dropping out samples from intermediate diffusion steps while preserving the noisy speech conditions, thereby forcing the model to rely more heavily on the conditional information during generation.
This combined approach enables the system to recover clean speech in as few as two steps, significantly reducing inconsistencies between the generated speech and its noisy condition.
\citet{shi2024diffusion} introduce a joint predictive-generative system that utilizes a shared encoder. 
This system incorporates the predictive model at both the initial and final diffusion steps. 
The initial fusion employs the predictive speech enhancement output to initialize the diffusion process, thereby improving convergence. 
At the final step, the system combines the decoded features from both predictive and generative decoders at the feature level, leveraging their complementary strengths to enhance overall speech enhancement performance.

\subsection{Source Separation} 
\noindent Audio source separation, which aims to isolate individual sound sources from complex mixtures, shares similarities with audio restoration in its problem definition. 
Consequently, this field has experienced comparable advancements in both predictive models~\citep{luo2019conv, defossez2019music, takahashi2020d3net, defossez2021hybrid, kilgour2022text, luo2023music} and generative models~\citep{postolache2023latent, zhu2022music, mariani2023multi, subakan2018generative, karchkhadze2024simultaneous, chen2023sepdiff, lutati2023separate}. 

One approach to source separation is to treat it as a mixture-conditioned source generation task.
For example,~\citet{chen2023sepdiff} introduced SepDiff for speech separation, which conceptualizes the task as mixture-conditioned generation. 
This method simply concatenates the mixture along the channel dimension of the input mel-spectrogram of individual sources, utilizing diffusion models to output individual speech sources with a fixed number of channels.
DiffSep~\citep{scheibler2023diffusion} adapts the OUVE process from SGMSE~\citep{richter2023speech} by restructuring the drift term to incorporate both noising and mixing operations at each step, while maintaining the same means for initial and target distributions.
Inspired by cold diffusion~\citep{bansal2024cold},~\citet{roglans2022diffusion,roglans23carnatic} propose an iterative approach for singing voice extraction where the initial vocal signal is deterministically corrupted by gradually introducing components from the full mixture.
Recent studies~\citep{mariani2023multi, karchkhadze2024simultaneous,postolache2024generalized} have explored unified frameworks that can perform both music source separation and conditional generation within a single model. 
These approaches utilize multi-source diffusion processes to model the joint distribution of all individual sources, enabling tasks like source imputation (generating compatible tracks for existing sources), unlike earlier methods that primarily focused on extracting individual sources from mixtures independently.

Another popular approach adopts a prediction-then-generation strategy. 
\citet{lutati2023separate} extended the theoretical upper bound for source separation to include generative models, indicating a potential signal-to-distortion Ratio (SDR) improvement of up to 3 dB compared to purely predictive methods. 
This approach first employs a predictive separation network to obtain initial estimates. 
These estimates are then refined using a pre-trained diffusion model.  
Subsequently, a shallow neural network, trained with a permutation-invariant loss, processes both the initial predictive estimates and the diffusion-refined estimates to produce mixing weights. 
The final separation is achieved by linearly combining both estimates in the spectral domain using these weights.
\citet{wang2024noise} proposed GeCo, which combines a deterministic predictor and generator. 
In GeCo, the drift and diffusion terms in the score SDE are modified based on the Brownian bridge SDE. 
This modification effectively removes noises and perceptually unnatural distortions introduced by the initial deterministic separator, resulting in robust improvements over the initial separation.

A key consideration in source separation is the definition of target sources.
For target speaker extraction~\citep{delcroix2018single}, there is a single target source in the mixture, while for music source separation and universal source separation (USS)~\citep{kavalerov2019universal}, the mixture can contain multiple target sources.
Both scenarios require the target sources to be predefined, often through training recordings of these target sources.
During inference, not only the mixture is input to the diffusion model, embeddings derived from the target sources or multi-hot source category encodings are also input to the diffusion model as layer modulation conditions, facilitating more effective target-conditional modeling~\citep{hai2024dpm,kamo2023target}.

In recent years, language-guided source separation (LASS) has emerged as a novel field within source separation~\citep{liu2022separate, kilgour2022text}. 
LASS distinguishes itself by defining target source categories through free-form text, allowing users to provide detailed, context-specific instructions. This approach enables models to separate sounds based on the unique characteristics described in textual queries.
Two notable implementations in this domain are FlowSep~\citep{yuan2024flowsep} and SoloAudio~\citep{wang2024solo}. 
FlowSep uses an FLAN-T5 encoder~\citep{chung2024scaling} to embed textual queries, which then inform a separation network based on rectified flow matching. 
In contrast, SoloAudio employs latent diffusion, aligning its latent space with the CLAP embedding space. 

In addition to their formulation as conditional generative models, both speech enhancement and source separation can be framed as inverse problems, particularly through the lens of Eq.~\eqref{eq:classifier}.
In this framework, the prior score term describes a trajectory towards the manifold of clean speech signals, while the likelihood score characterizes a path to the manifold of signals consistent with the observed degraded data.
During sampling, these two components work jointly to guide the trajectory towards solutions that satisfy both constraints simultaneously: The prior score keeps signal in distribution with clean training data, while the likelihood score ensures consistency with observations~\citep{lemercier2024diffusion}.
This approach effectively reconciles the estimated clean speech with both the prior expectations of speech characteristics and the constraints imposed by the observed degraded signal.
\citet{jayaram2020source} introduced a generator-agnostic approach called Bayesian annealed signal source separation (BASIS), a method prior to the widespread of DDPMs that shares conceptual similarities with guided sampling approaches introduced in Sec.~\ref{sec:sg}. 
BASIS trains a prior distribution on clean source data using either noise conditional score networks (NCSN)\citep{song2019generative} or Glow~\citep{kingma2018glow}. 
The method formulates source separation as sampling from a posterior distribution, combining the learned prior with a likelihood term that approximates the mixture constraint using a Gaussian distribution with parameterized variance.
For efficient inference, BASIS employs Langevin dynamics with gradually annealed noise levels, allowing sampling from the posterior without explicitly computing intractable partition functions. 
This formulation using generative priors as signal models inspired several subsequent works employing various generators~\citep{postolache2023latent,zhu2022music,jayaram2021parallel}. 
Modern diffusion-based generators typically adopt diffusion posterior sampling (DPS)~\citep{chung2023diffusion}, which utilizes more sophisticated techniques to approximate the likelihood through clean data estimation derived from noisy versions.
The versatility of DPS has led to its application in a wide range of audio tasks, using task-specific degradation operator in the likelihood term. 
These tasks include dereverberation~\citep{lemercier2023diffusion}, source separation~\citep{iashchenko2023undiff}, inpainting~\citep{moliner2024diffusion}, and bandwidth extension~\citep{moliner2023solving,moliner2023blind}.

\chapter{Code Implementation}
\label{sec:code}

We provide a comprehensive codebase~\footnote{\url{https://github.com/gzhu06/AudioDiffuser}} that implements the reviewed diffusion modeling approaches into a modular framework illustrated in Fig.~\ref{fig:diffsummary}. 
Our implementation is based on the EDM framework that supports a range of diffusion models and techniques discussed throughout this review, enabling experimentation across different configurations and audio applications.
Following the architecture presented in Fig.~\ref{fig:diffsummary}, we developed our codebase using PyTorch Lightning~\citep{falcon2019lightning} and Hydra~\citep{Yadan2019Hydra}, allowing each component to be independently controlled through configuration YAML files. 
Building upon the {\fontfamily{qcr}\selectfont audio-diffusion-pytorch} library\footnote{\url{https://github.com/archinetai/audio-diffusion-pytorch}, release v0.0.94}, our implementation realizes four primary modular components that can be independently configured and combined:

\section{Training module}
We implement multiple training objectives including noise prediction, velocity prediction, denoising score matching, and flow matching (rectified flow) formulations discussed in Sec.~\ref{sec:train_obj}. 
The corresponding training noise schedule module supports various distributions including uniform, cosine-uniform, sigmoid-uniform, log-normal, and logit-normal schedules as reviewed in Sec.~\ref{sec:trainnoise}.

\section{Sampling module}
Our deterministic sampler implementations include first-order Euler's method, second-order Heun's method, and DPM-Solver variants from Sec.~\ref{sec:sampling}, integrated from multiple sources including the official EDM codebase\footnote{\url{https://github.com/NVlabs/edm}}, DPM~\citep{lu2022dpm,lu2022dpmpp} and UniPC~\citep{zhao2024unipc} repositories\footnote{\url{https://github.com/LuChengTHU/dpm-solver,https://github.com/wl-zhao/UniPC}}, and the k-diffusion framework\footnote{\url{https://github.com/crowsonkb/k-diffusion}}.
The sampling noise schedule module provides linear, quadratic, cosine LogSNR, linear LogSNR, log-linear, and polynomial (EDM) scheduling options. 
To address compatibility issues across different framework implementations from multiple sources, we include a cross-compatibility mechanism that enables models trained with different objectives to work seamlessly with various inference samplers. 
This mechanism employs automatic rescaling functions that adapt both noisy inputs and noise levels to match each network's native training distribution, with customized adaptations for Euler-based methods (Euler, Heun) and exponential integrator-based approaches (DPM-solver, UniPC).
Additionally, our adopted Lightning-Hydra template~\footnote{\url{https://github.com/ashleve/lightning-hydra-template}} enables hyperparameter grid search, providing efficient exploration of sampling parameters to optimize inference-time performance.

\section{Model architecture module}
The application conditioning component supports multiple conditioning mechanisms including AdaLN (FiLM), cross-attention, sequence concatenation, and channel concatenation as discussed in Sec.~\ref{sec:cond}. The network backbone module accommodates various architectures including WaveNet, U-Net, and Transformers reviewed in Sec.~\ref{sec:ad}.

\section{Application module}
Following Sec.~\ref{sec:app}, our framework supports diverse conditional generation tasks including text-based generation (with labels, text, speech transcripts) and audio enhancement (with noisy audio input). 
For each application category, we provide specialized data loading, preprocessing, and conditioning implementations, including text encoding modules for text-based generation and noise augmentation pipelines for enhancement tasks.

\chapter{Experiment}



In this section, we evaluate the modular framework through experiments on both unconditional and conditional audio generation tasks, utilizing the comprehensive codebase described in Sec.~\ref{sec:code}.
Our evaluation has two primary objectives that leverage the framework's modular design. 
First, we assess the core generative modeling process discussed in Sec.~\ref{sec:main} through an unconditional audio generation experiment that systematically compares different training objectives, network architectures, and sampling approaches.
This comparative study utilizes the training module's multiple objective implementations and the sampling module's deterministic samplers from Sections~\ref{sec:train} and~\ref{sec:sampling}.
Second, we validate the model architecture module's flexible conditioning mechanisms discussed in Sec.~\ref{sec:app} through conditional audio generation experiments that cover two representative tasks: speech enhancement (employing in-context conditioning with noisy audio input) and TTS (utilizing cross-attention for text transcription input and modulation for speaker embeddings).
These tasks were selected for their accessible public datasets and their representation of diverse conditioning approaches discussed in Section~\ref{sec:ad}, both supported by our application module's specialized implementations.
Together, these experiments demonstrate the practical applicability of our summarized modular framework and evaluate the effectiveness of our hyperparameter optimization strategies enabled by the Lightning-Hydra grid search capabilities in real-world generation scenarios.

\section{Unconditional Audio Generation}
\label{sec:ag}
\subsection{Experimental Setup} 
For unconditional audio generation task, we use the speech command 09 (SC09) dataset~\citep{warden2018sc09}, which contains spoken digits.
We evaluate models trained with various diffusion objectives and noise distributions, employing different samplers with their corresponding noise schedulers.
In Table~\ref{tab:specifics}, we present the training parameters and configurations representing the most commonly used diffusion objective setups in current practices.
Specifically, we formulate VE and VP objectives within the EDM framework, alongside the v-diffusion approach described in simple diffusion~\citep{hoogeboom2023simple} and the rectified flow methodology from stable diffusion~\citep{esser2024scaling}.
The training configurations outlined in Table~\ref{tab:specifics}, including loss weighting, training and inference noise schedules, are based on parameters that have proven effective for image generation.
However, due to computational constraints, we adopt these parameters without extensive hyperparameter optimization for audio data. 
As demonstrated by~\citet{esser2024scaling}, determining truly optimal noise distributions for each objective requires significant computational resources, which we defer to future work.

\begin{sidewaystable}
\centering
\Large
\resizebox{\textwidth}{!}{
\begin{tabular}{lccccc}
\toprule
& VE~\citep{karras2022elucidating} & VP~\citep{karras2022elucidating} & V-obj.~\citep{hoogeboom2023simple} & EDM~\citep{karras2022elucidating} &RF~\citep{esser2024scaling}  \\
\midrule
\multicolumn{5}{l}{\textbf{Training (Sec.\ref{sec:train})}} \\
Training objective & $F_\theta(c_{\text{in}}(\sigma)\mathbf{x},\ln(\frac{1}{2}\sigma))$ & $F_\theta(c_{\text{in}}(\sigma)\mathbf{x},(M-1)\cdot t)$ & $v_\theta(\mathbf{x},\gamma)$&$F_\theta(c_{\text{in}}(\sigma)\mathbf{x},\frac{1}{4}\ln(\sigma))$ &$v_\theta(\mathbf{x},\sigma)$\\
 &   & &\\
Noise distribution & 
$\sigma\sim\sigma_{\text{min}}(\frac{\sigma_{\text{max}}}{\sigma_{\text{min}}})^{\mathcal{U}(0,1)}$ & $t \sim \mathcal{U}(\epsilon_t, 1)$ & $t \sim \mathcal{U}(0, 1)$& $\sigma \sim e^{\mathcal{N}(P_{\text{mean}}, P^2_{\text{std}})}$ & $\sigma \sim \text{logistic} (\mathcal{N}(P_{\text{mean}}, P^2_{\text{std}}))$\\
\midrule
\multicolumn{5}{l}{\textbf{Sampling (Sec.\ref{sec:sampling})}} \\
Time steps $t_{i<N}$ & 
$\sigma^2_{\text{max}}(\frac{\sigma^2_{\text{min}}}{\sigma^2_{\text{max}}})^{\frac{i}{N-1}}$ & $1 + \frac{i}{N-1}(\epsilon_s - 1)$ & 
$1 - \frac{i}{N-1}$ & $(\sigma_{\max}^{\frac{1}{\rho}}+\frac{i}{N-1}(\sigma_{\min}^{\frac{1}{\rho}}-\sigma_{\max}^{\frac{1}{\rho}}))^{\rho}$ &$1 - \frac{i}{N-1}$\\
Schedule $\sigma(t)$ & $\sqrt{t}$& $\sqrt{e^{\frac{1}{2}\beta_dt^2+\beta_{\text{min}}t}-1}$  & $\sqrt{\text{sigmoid}(-\gamma(t))}$ & $t$ & $t$\\
&&&$\gamma(t)=-2\log\tan(\text{lerp}(t_{\min},t_{\max},t))+s$\\
\midrule
\multicolumn{5}{l}{\textbf{Parameters}} \\
& $\sigma_{\text{min}} = 0.02$& $\beta_d = 19.9, \beta_{\text{min}} = 0.1, M=1000$  
& $t_{\text{min}}=\tan^{-1}(e^{-15/2})$ & $\sigma_{\text{min}} = 0.002, \sigma_{\text{max}} = 80, \rho = 7$ &$P_{\text{mean}}=0$\\
& $\sigma_{\text{max}} = 100$ & $\epsilon_s = 10^{-3}, \epsilon_t = 10^{-5}$ &  $t_{\text{max}}=\tan^{-1}(e^{15/2}),s=0$ & $P_{\text{mean}}=-1.2, P_{\text{std}}=1.2$ & $P_{\text{std}}=1$\\
\bottomrule
\end{tabular}}
\caption{Comparison of five diffusion formulations for SC09 benchmark. 
For each formulation, we detail the training objective $F_{\theta}$ or $v_{\theta}$, noise distribution, time step sequence $t_{i<N}$, noise schedule $\sigma(t)$, and their associated parameters.}
\label{tab:specifics}
\end{sidewaystable}

To address our second experimental goal of providing insights on different diffusion approaches, we evaluate various diffusion objectives using the same U-Net architecture with identical hyperparameters.
This controlled comparison isolates the impact of the diffusion objective itself while maintaining all other variables constant. 
All models are trained for 1000 epochs with a batch size of 32, using the AdamW optimizer with a learning rate of 5e-4 with a decay rate of 0.01 and an exponential moving average (EMA) decay rate ($\beta$) of 0.9999.
For our neural network backbone, we implement an efficient U-Net architecture adapted from the open-source Imagen implementation\footnote{\url{https://github.com/lucidrains/imagen-pytorch}}. 
This architecture is optimized for memory efficiency and fast convergence. 
For the network input, we use complex spectrogram with window size of 510 samples and hop size of 256 samples. 
The input consists of 2 channels representing real and imaginary components, with an initial embedding dimension of 128. 
This initial dimension serves as the reference value from which subsequent U-Net block channel dimensions are derived as multiples (e.g., 128, 256, 512) throughout the network architecture.
Each downsampling/upsampling block incorporates 2 ResNet blocks, each containing a self-attention layer with 2 attention heads. 
The complete model comprises 46.9 million trainable parameters. 
We condition the network on the logarithm of noise levels $\log(\sigma)$.

We evaluate a comprehensive range of deterministic samplers across different diffusion formulations to demonstrate both the flexibility of our evaluated framework and provide insights into sampler effectiveness for audio generation. 
By implementing first-order Euler's method, second-order Heun's method~\citep{karras2022elucidating}, EI-based DPM~\citep{lu2022dpm,lu2022dpmpp}, and UniPC~\citep{zhao2024unipc} samplers, we showcase our codebase's ability to support diverse sampling strategies while enabling direct performance comparisons. 
This evaluation serves our experimental goals by validating that our framework can successfully integrate multiple sampling approaches and establishing empirical evidence for which training objectives and samplers achieve better quality-efficiency tradeoffs in audio tasks.


Following~\citet{goel2022sashimi,kong2020diffwave}, we evaluate the model's unconditional generation performance using four standard metrics: Frechet inception distance (FID), inception score (IS), modified inception score (mIS), and AM Score.

\subsection{Experimental Results}
\begin{figure*}[!t]
\centering
\includegraphics[width=\textwidth]{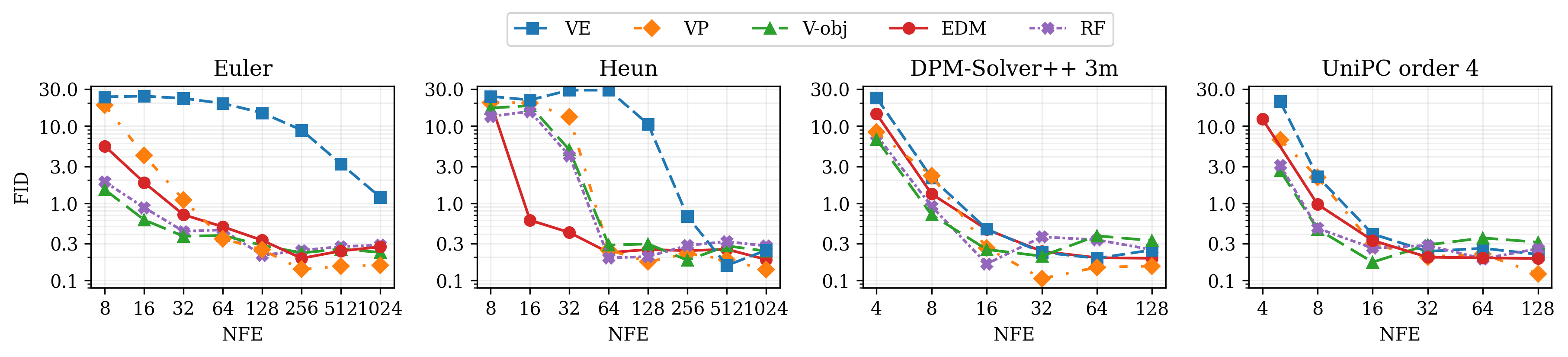}
\caption{FID scores versus NFE for unconditional audio generation on SC09 across four sampling methods.
Each line represents a different diffusion formulation. Lower FID scores indicate better generation quality. Both axes use a logarithmic scale.
}
\label{fig:fidvsstep}
\end{figure*}

We evaluate four deterministic samplers (Euler, Heun, DPM-Solver++ 3m, and UniPC order 4) across abovementioned five diffusion parameterizations (VE, VP, V-obj., EDM and RF). Fig.~\ref{fig:fidvsstep} shows the relationship between FID scores and NFE for these configuration combinations. 
It can be seen that the FID score generally \gz{decreases} with NFE until convergence for all configurations, although with different convergence rates.
The Euler sampler requires the highest NFE to achieve the optimal FID scores.
Higher-order methods (Heun, DPM-Solver++, and UniPC) achieve comparable or superior FID scores with significantly reduced NFE, with DPM-Solver++ and UniPC demonstrating effective performance using as few as 16 NFEs, aligning with the efficiency claims in~\citet{lu2022dpm} and~\citet{zhao2024unipc}. 
Among diffusion parameterizations, the v-diffusion (red) and RF (purple) parameterizations show similar rapid convergence patterns, confirming the theoretical advantages described in~\citet{liu2023flow, zheng2023improved}.
EDM (blue) exhibits rapid initial convergence with all samplers, while VE (orange) consistently requires higher NFE to achieve comparable FID scores. 
VP (green) demonstrates competitive FID scores, particularly at higher NFE. 
These comparative findings regarding the performance characteristics of EDM, VE, and VP in unconditional audio generation align with previous observations in image generation by~\citet{karras2022elucidating}, suggesting consistent behavior across different data modalities.

We conduct a comprehensive grid search over all sampler hyperparameters to optimize FID scores for each diffusion parameterization, implementing this optimization process using the Lightning and Hydra frameworks.  
After identifying the optimal configuration based on FID performance, we evaluate these configurations using additional metrics, with results presented in Table~\ref{tab:gridsearch}.
Within our search space, the optimal configurations achieve comparable FID scores at varying NFEs.
Particularly, VP with DPM++ 3m achieves the best FID score (0.106) at the lowest computational cost (32 NFEs), demonstrating superior efficiency. In contrast, EDM with UniPC-2 achieves the highest perceptual quality metrics (IS: 7.98, mIS: 257.6), though at a higher computational cost (128 NFEs).

\begin{table}[t]
\centering
\resizebox{0.8\textwidth}{!}{
\begin{tabular}{l c c c c c c}
\toprule
& Sampler & NFE ($\downarrow$) & FID ($\downarrow$) & IS ($\uparrow$)& mIS ($\uparrow$) & AM ($\downarrow$)\\
\midrule
VE & Heun & 256& 0.156 & 7.05 & 149.3 &0.345\\
VP & DPM++3m& \bf{32} & \bf{0.106} & 7.93  & 228.2& 0.241\\
v-diffusion& DPM 3m & 64 & 0.155 & 7.96 & 237.1 &0.234\\
EDM & UniPC-2 & 128 & 0.141  & \bf{7.98} & \bf{257.6} & \bf{0.227}\\
RF & UniPC-3&64&0.117& 7.82 & 218.8 & 0.248\\
\bottomrule
\end{tabular}}
\caption{Best sampler configurations and performance metrics for different diffusion formulations on the SC09 unconditional audio generation benchmark.
Results show the best FID scores achieved for each method after a grid search on several sampler hyperparameters. These hyperparameters and their ranges are: sampling steps across all methods, solver orders (1-3) and prediction objectives ($x_0$ or $\epsilon$) for DPM-solver with single/multi-step variants, and solver orders (2-4) with prediction objectives ($x_0$ or $\epsilon$) for UniPC. 
Arrows indicate whether lower or higher values are better for each metric.}
\label{tab:gridsearch}
\end{table}

\section{Speech Enhancement}
\subsection{Experimental Setup} 
For speech enhancement model training, we constructed a large-scale dataset of paired clean and noisy speech following the URGENT 2024 challenge preprocessing recipe~\citep{zhang2024urgent}.
To evaluate model performance, we utilized two test sets as proposed in~\citet{Scheibler2024universepp}: the Voicebank + DEMAND (VB)~\citep{botinhao2016speech} test split, and its modified version with low-pass filtering at 4 kHz cutoff frequency (VB-BWE) to assess the model's bandwidth extension capabilities.

The training dataset consisted of synthetic noisy speech generated by mixing clean speech, noise signals, and room impulse responses (RIRs). 
For clean speech, we utilized three public corpora: the LibriVox dataset from the DNS5 challenge~\citep{dubey2024icassp}, the English portion of CommonVoice 11.0~\citep{ardila2020common}, and LibriTTS~\citep{zen2019libritts}. 
To prevent potential data leakage during evaluation, we excluded the VCTK portion from our training set.
The noise and RIR components were sourced from the combined Audioset and FreeSound noise datasets from the DNS5 challenge, and the WHAM! noise dataset~\citep{Wichern2019WHAM}. 
Following URGENT challenge, to evaluate our model's capability across different bandwidths, we resampled the input noisy speech signals to various sampling rates ranging from 8 kHz to 48 kHz, while maintaining the target clean speech at the same sampling rate post-enhancement.
During training, all input signals were uniformly upsampled to 48 kHz, and we utilized power-compressed complex spectrograms as feature inputs.

We extended our efficient U-Net backbone (detailed in Sec.~\ref{sec:ag}) to conditional generation for speech enhancement application. 
The primary architectural modification involves introducing an auxiliary conditioner network that mirrors the main U-Net encoder's parameters.  
This conditioner network processes the noisy speech complex spectrogram and extracts feature maps at each encoder block. 
These feature maps are then combined with the corresponding encoder outputs of the main score network through feature fusion. 
While various feature fusion approaches exist, such as channel concatenation or applying additional convolutional blocks, we opted for simple element-wise addition for its effectiveness and computational efficiency.
For model training, we maintained consistency with our audio generation approach by employing the same EDM objective and feature parameters described in Sec.~\ref{sec:ag}. 
Similarly, we utilized the DPM-solver for generation, leveraging its computational advantages. 
We optimized the model's inference performance through sampler hyperparameter tuning, employing the Deep noise suppression mean opinion score (DNSMOS)~\citep{reddy2021dnsmos} as our primary evaluation metric. 
DNSMOS was chosen for its reliability in assessing speech quality without requiring reference signals, making it particularly suitable for real-world applications. 
We focused our optimization efforts on the DPM-solver using the same parameter set detailed in Sec.~\ref{sec:ag}.
Through this grid search over these hyperparameters, we identified an optimal configuration that maximized DNSMOS scores across our evaluation sets while maintaining efficiency with only 8 NFEs.
This optimized sampling configuration, along with its corresponding training parameters, defines our implemented `EDM Enhancer' model.

For model comparison, we include both predictive and generative approaches from~\citet{Scheibler2024universepp}. 
We also trained a state-of-the-art predictive model, TFGridNet\citep{wang2023tf}, on our constructed dataset using a multi-resolution L1 spectral reconstruction objective.
We adopt the comprehensive evaluation suite proposed in~\citet{Scheibler2024universepp}, incorporating a diverse set of speech enhancement metrics~\citep{pirklbauer2023evaluation}.
To assess waveform reconstruction quality, we employed PESQ-WB~\citep{rix2001perceptual} and log-spectral distance (LSD), while speech intelligibility was measured using ESTOI~\citep{jensen2016algorithm}. 
For linguistic content preservation, we utilized Levenshtein phoneme similarity (LPS), which quantifies phoneme accuracy between enhanced and clean speech using a phoneme recognition model~\citep{pirklbauer2023evaluation}. 
Speech naturalness was evaluated using reference-free DNSMOS~\citep{reddy2021dnsmos}.

\subsection{Experimental Results}

\begin{table*}[h!]
\centering
\setlength{\tabcolsep}{2pt}
\resizebox{\textwidth}{!}
{
    \begin{tabular}{c c cccccccccc}
    \toprule

\multirow{2}{*}{Model} & \multirow{2}{*}{Type} & \multicolumn{5}{c}{VB} & \multicolumn{5}{c}{VB-BWE} \\[-2ex]
    & & \multicolumn{5}{c}{\leaders\hrule height 0.4pt\hfill\kern0pt} & \multicolumn{5}{c}{\leaders\hrule height 0.4pt\hfill\kern0pt} \\

          &  & PESQ-WB $\uparrow$ & ESTOI $\uparrow$ & LSD (dB) $\downarrow$ & LPS (\%) $\uparrow$ & DNSMOS $\uparrow$ & PESQ-WB $\uparrow$ & ESTOI $\uparrow$ & LSD (dB) $\downarrow$ & LPS (\%) $\uparrow$ & DNSMOS $\uparrow$ \\ 
    \midrule
    Unprocessed            &                       & 1.97 & 0.787 & 10.3 & 93.2 & 2.68 & 2.05 & 0.786 & 10.0 & 91.8  & 2.70 \\
    \midrule
    BSRNN*                  & Pred            & 2.53 & 0.848 & 6.0  & 94.5 & 3.00 & 2.58 & 0.838 & 9.0  & 89.9  & 2.92 \\
    TFGrid                 & Pred            & 2.64 & \textbf{0.865} & \textbf{5.3}  & \textbf{94.8} & 3.16 & 2.58 & \textbf{0.864} & \textbf{6.9}  & \textbf{93.8}  & 3.15 \\
    \midrule
    StoRM* & Pred + Gen & 2.41    & 0.845 & 7.0      & 94.6     & 3.07   & 1.62    & 0.812 & 19.1     & 88.9    & 2.79   \\
    UNIVERSE*               & Gen             & 2.55 & 0.784 & 9.7  & 87.9 & 3.12 & 2.36 & 0.768 & 10.3 & 82.7 & 3.11 \\
    UNIVERSE++*             & Gen             & \textbf{2.95} & 0.860 & 6.2  & 94.0  & \textbf{3.20} & \textbf{2.72} & 0.848 & 9.1  & 90.3 & \textbf{3.19} \\
    EDM Enhancer (Ours)                   & Gen             & 2.66 & 0.855 & 6.4  & 94.0 & 3.18 & 2.71 & 0.856 & 10.7 & 93.3 & 3.16 \\
    \bottomrule
    \end{tabular}
}
\caption{Results on two benchmark datasets: VoiceBank+DEMAND (VB) and its bandwidth-extension variant VB-BWE (4kHz $\to$ 24kHz). 
Methods with * show metrics reported from~\citet{Scheibler2024universepp}.
Due to potential differences in ASR model configurations, word error rate (WER) results may not be directly comparable with those in \citet{Scheibler2024universepp}, so we remove this comparison.}
\label{tab:se_eval}
\end{table*}

Table~\ref{tab:se_eval} presents evaluation results on the two benchmarks against baseline models' performance. 
The evaluation includes both predictive and generative approaches, indicating advantages from different aspects.
On the VB benchmark, TFGridNet, a predictive approach, demonstrates superior performance across multiple objective metrics. 
Specifically, it achieves notable improvements in speech intelligibility (ESTOI), spectral distortion measures (LSD), and speech content accuracy (LPS). 
The model's exceptional LSD scores can be attributed to its reconstruction-focused training objective, which emphasizes precise signal reproduction.

In contrast, generative methods exhibit different strengths. 
While TFGridNet shows superior performance in speech content fidelity, diffusion based EDM enhancer achieves comparable performance with TFGridNet in terms of perceptual quality metrics.
Notebly, UNIVERSE++ achieves superior results in perceptual quality metrics (PESQ-WB and DNSMOS).
This superior perceptual performance can be attributed to UNIVERSE++'s auxiliary GAN-based loss, which promotes natural-sounding output. 
These findings align with previous studies~\citep{pirklbauer2023evaluation,Scheibler2024universepp}, which establish that predictive methods excel in preserving sample-wise content, while generative approaches tend to produce more naturalistic audio output.
The VB-BWE benchmark results indicate similar trends. 
TFGridNet maintains its strong performance in ESTOI, LSD, and LPS metrics. 
Through informal subjective evaluation on the band-limited speech signals, we observed that TFGridNet effectively preserves content below the cutoff frequency but introduces artifacts in bandwidth extension regions.  
Consistent with the VB benchmark results, UNIVERSE++ and our method achieve the highest and second-highest PESQ-WB and DNSMOS scores, respectively.
Our method, while not leading in either perceptual metrics (compared to UNIVERSE++) or sample-wise metrics (compared to TFGridNet), demonstrates competitive performance across both domains. 
Notably, these results are achieved using solely a diffusion training objective without auxiliary losses, highlighting the effectiveness of our approach.

The above results validate several key aspects of our reviewed framework. 
First, they demonstrate the effectiveness of in-context conditioning mechanisms discussed in Sec.~\ref{sec:ad}, showing that our framework can effectively incorporate auxiliary conditions through feature fusion approaches. 
Second, the competitive performance achieved with only 8 NFEs validates our sampler hyperparameter tuning strategy. 
Third, the comparison between generative and predictive methods also provides practical insights into the trade-offs between content fidelity and perceptual quality that are inherent to diffusion-based approaches. 

\section{Text-to-Speech}
\subsection{Experimental Setup}
We developed our TTS generation model using a curated subset of the LibriVox dataset\footnote{\url{https://librivox.org/}}. 
The data preprocessing pipeline consisted of two main stages: First, we applied voice activity detection (VAD) using WhisperX package~\citep{bain2022whisperx} to segment full reading episodes into shorter utterances ranging from 4 to 10 seconds. 
Second, we transcribed these segmented utterances through Whisper-large-v2 model to generate accurate text transcriptions, creating aligned audio-text pairs for training.
The final training corpus consisted of 5 million pieces of read speech utterances paired with text transcripts, totaling 8,500 hours.
For evaluation, we chose not to use the LibriSpeech test set~\citep{panayotov2015librispeech}, despite its widespread adoption in the field, to avoid potential data contamination since our training data was directly derived from LibriVox.
Instead, we evaluated synthesis performance on a subset of VCTK corpus, constructed by randomly sampling 50 utterances per speaker. 
To maintain consistency with our training distribution, we selected utterances with word counts matching the distribution observed in our training split.

Our TTS system adopts a latent diffusion framework, which effectively separates semantic and perceptual information processing. 
This separation is particularly beneficial for generating long-form speech sequences, as it enables efficient processing in a compressed latent space.
For the speech autoencoder component, we modify the existing DAC autoencoder~\citep{kumar2023high} by converting its discretized tokens into a continuous representation.
Our initial attempts on training EDM diffusion models directly on the continuous latents before the VQ-bottleneck were unsuccessful, failing to generate meaningful results due to the high dimensionality of the latent space.
This aligns with observations by~\citet{esser2024scaling}, suggesting that such high-dimensional spaces likely require substantially larger diffusion backbones. 
To address this limitation, we develop a mini-VAE that reduces the dimensionality of the original DAC latent space while maintaining the time compression rate. 
The architecture consists of four convolutional encoder/decoder blocks that progressively reduce the dimension from 1024 to 32. 
We opt for a simplified training approach by freezing both the encoder and decoder, without incorporating GAN loss for perceptual improvement.
During training, we standardize input processing by padding audio sequences to 10 seconds and apply masking to reduce the impact of silence segments~\citep{gao2023e3}. 
At inference time, we post-process the generated audio through an additional energy-based VAD post-processing step.

For the diffusion network backbone, we employ DiT due to its demonstrated scalability on large-scale datasets.
We term our resulting system E3-DiT, as it builds upon the non-autoregressive design principles of E3-TTS~\citep{gao2023e3}. 
The system relies solely on language embeddings from transcriptions and speaker embeddings, eliminating the need for external aligners or prosody modeling components.
Following SoundStorm~\citep{borsos2023soundstorm}, we extract language embeddings from raw text transcriptions using byT5~\citep{xue2022byt5}. 
While our implementation uses these pre-trained embeddings, the architecture can accommodate learnable text encoders that accept either tokenized text or byte-level characters as input when memory constraints are not a concern. 
For the conditioning mechanism, rather than implementing cross-attention, we adopt the approach from PlayGroundv3~\citep{liu2024playground} where text and audio tokens are treated as a unified sequence and concatenated at each self-attention layer. 
We then apply rotary positional embedding (RoPE)~\citep{su2021roformer} to this combined sequence.
For speaker conditioning, we utilize the ECAPA-TDNN model~\citep{desplanques20ecapa} to extract speaker embeddings, chosen for its state-of-the-art performance in speaker recognition. 
These speaker embeddings are incorporated into the model through adaptive layer normalization.
During inference, we evaluate the influence of reference speaker embeddings using both 5-second and 20-second audio segments.

Prior to feeding the latents into the network, we standardize them and set the data standard deviation $\sigma_{\text{data}}$ to 0.5.
Our transformer backbone consists of 16 layers with a hidden size of 1536 and 16 attention heads, processing sequences of length 806 with a patch size of [1, 4]. 
For text processing, we utilize the byT5-base model to extract embeddings with a dimension of 1536, accommodating text sequences up to 120 tokens in length. 
The noise distribution for training follows a log-normal distribution (mean=-0.4, std=1.2), consistent with the latent diffusion parameterization used in EDM2~\citep{karras2023analyzing}.
The training process incorporates CFG with a dropout rate of 0.1.
We optimize the model using Adam with a learning rate of 1e-4 with a batch size of 24 over 1 million iterations.

We evaluate our model's performance across three critical dimensions: speech intelligibility, voice preservation, and audio quality, following the evaluation framework established by Soundstorm.
For speech intelligibility, we utilize the NeMo package~\citep{kuchaiev2019nemo} to compute both word error rate (WER) and character error rate (CER) on the generated speech. 
Voice preservation is assessed by calculating the cosine similarity between speaker embeddings extracted from the reference speaker and those from the generated audio.
To evaluate audio quality, we employ two metrics. 
We compute FAD using the Trill model~\citep{shor2020towards}, a non-semantic speech representation that has been shown to effectively capture vocal quality characteristics~\citep{huang2023noise2music}. 
Additionally, we estimate MOS using Torchaudio-Squim to assess the perceived quality of the generated samples. 
Similarly, during the optimization of inference parameters, we primarily focus on the reference-free metrics -- speaker similarity, FAD, and MOS -- to tune the DPM-solver hyperparameters.
The searched hyparameters are conducted only for DPM-solver for simplicity, besides the above parameters, we also include CFG scales ranging from 6 to 15 with a step size of 3.
After this round of search, the resulting sampler requires approximately 32 NFEs.

\subsection{Experimental Result} 
Table~\ref{tab:tts_eval} presents our evaluation results on the VCTK benchmark, comparing our E3-DiT variants against oracle autoencoder reconstruction performance.
The DAC-miniVAE reconstruction baseline demonstrates superior performance across all metrics, establishing a strong upper bound for the latent diffusion approach.

Our framework evaluation reveals several key insights regarding practical model design choices. 
First, the comparison between different speaker embedding lengths (5-second vs. 20-second) demonstrates the impact of conditioning signal quality on generation performance. 
Models using 20-second speaker embeddings consistently outperform their 5-second counterparts in speaker similarity (0.74 vs. 0.71) and speech intelligibility, validating the importance of sufficient conditioning.
Second, our sampler optimization experiments across three objectives (speaker similarity, FAD, and MOS) provide practical insights into hyperparameter tuning strategies. 
Speaker similarity optimization consistently produced better WER/CER scores across both embedding lengths, while FAD and MOS optimization yielded comparable results, suggesting that these metrics capture similar perceptual aspects.

These TTS results provide several key insights that validate our reviewed framework. 
First, the successful implementation of multiple conditioning mechanisms, text conditioning through sequence concatenation and speaker conditioning via adaptive layer normalization, demonstrates the flexibility of our framework's conditioning approaches from Sec.~\ref{sec:ad}.
Particularly, the concatenation-based approach from PlayGroundv3 proves to be an effective substitute for the cross-attention mechanisms discussed in Section~\ref{sec:ad}.
Second, both speech enhancement and TTS experiments confirm that our framework achieves strong performance using solely the EDM training objective without requiring auxiliary losses, demonstrating the robustness of the diffusion formulation for conditional audio generation.
Finally, the comprehensive sampler optimization across multiple objectives validates the framework's adaptability to different generation priorities, confirming its practical utility for real-world applications with varying performance requirements.


\begin{table}[t!]
\centering
\resizebox{\columnwidth}{!}{
    \begin{tabular}{c c cc ccc}
    \toprule
    Model& Grid Search Obj. & WER $\downarrow$ & \textbf{CER $\downarrow$} & Spkr. Similarity $\uparrow$ & FAD $\downarrow$& MOS $\uparrow$  \\ 
    \midrule
    DAC-miniVAE-recon. &-& 0.49&0.30 & 0.742 & 0.131& 4.39\\
    \midrule
    \multicolumn{7}{l}{\textit{Ref. Spkr. duration: 5s}} \\
    E3-DiT-1 & Spkr. Similarity& 5.34 & 3.09 & 0.709 & 0.298 &4.21\\
    E3-DiT-2 & FAD&5.83 & 3.40 & 0.707 & 0.296 & 4.22\\
    E3-DiT-3 & MOS&5.73 & 3.29 & 0.707 & 0.297 & 4.23\\
    \midrule
    \multicolumn{7}{l}{\textit{Ref. Spkr. duration: 20s}} \\
    E3-DiT-4 & Spkr. Similarity& \textbf{4.86} & \textbf{2.77} & 0.741 & 0.297 &4.14\\
    E3-DiT-5 & FAD&5.14 & 3.00 & 0.739 & 0.302 & 4.13\\
    E3-DiT-6 & MOS& 5.08 & 2.98 & 0.739 & 0.298 & 4.14\\
    \bottomrule
    \end{tabular}
}
\caption{Comparative performance analysis of TTS models on the VCTK dataset. 
The DAC-miniVAE reconstruction serves as an upper-bound reference, representing ideal autoencoder performance. 
E3-DiT variants are grouped by speaker embedding context duration (5s vs 20s) and optimized using different reference-free objectives.}
\label{tab:tts_eval}
\end{table}

\chapter{Conclusions}
In this paper, we presented a comprehensive review of score-based generative models for audio applications, systematically examining diffusion model design through the unifying lens of the EDM framework.
We bridged the gap between theoretical foundations and practical implementations by detailing the components of diffusion models—from training objectives and noise distributions to sampling techniques and network architectures. 
Our review of audio applications demonstrated how these models have improved text-to-audio generation, speech synthesis, and audio restoration through various conditioning mechanisms. 
To facilitate reproducibility and accelerate research in this rapidly evolving field, we introduced an open-source codebase implementing the key components discussed in our review. 
Through our experimental case studies in unconditional audio generation, speech enhancement, and text-to-speech synthesis, we validated the effectiveness and flexibility of our modular framework. 
As score-based generative models continue to advance, we anticipate further innovations in training objectives, efficient stochastic sampling, conditioning techniques, and architectural designs that will expand the capabilities and applications of these powerful models across the audio domain.

\begin{acknowledgements}
We thank valuable discussions with Jordan Darefsky throughout the paper writing and codebase development.
We also thank Zachery Novak from UCSD and Jonah Casebeer, Prem Seetharaman, Rithesh Kumar and Nick Bryan from Adobe Research for the discussions on diffusion models.
\end{acknowledgements}

\backmatter  

\printbibliography

\end{document}